\documentclass[aps,prl,showpacs,preprintnumbers,amsmath,amssymb,groupedaddress]{revtex4}
\usepackage[dvips]{graphicx}
\usepackage{dcolumn}

\begin{document}

\title{Angular dependence of antiferromagnetic order induced by paramagnetism 
in $d$-wave superconductor}

\author{Ken-ichi Hosoya and Ryusuke Ikeda}

\affiliation{
Department of Physics, Graduate School of Science, Kyoto University, Kyoto 606-8502, Japan
}

\date{\today}

%less than 10 lines 

\begin{abstract} 
Antiferromagnetic (AFM) order and a spatial order peculiar to Fulde-Ferrell-Larkin-Ovchinnikov (FFLO) states, previously indicated in the quasi two-dimensional $d$-wave superconductors CeCoIn$_5$ with strong paramagnetic pair breaking (PPB) in a magnetic-field parallel to the basal plane, are considered in the field configurations tilted from the basal plane within an approach assuming that the wavelength of the FFLO modulation is relatively long. It is demonstrated that, with increasing the tilt angle, both the AFM and FFLO orders are gradually suppressed, and that disappearance of the AFM order in zero temperature limit occurs at a lower angle than that of the FFLO state. Consequently, a {\it nonmagnetic} FFLO-ordered high field SC phase is realized in an intermediate range of the tilt angle even at low enough temperatures. As the perpendicular field configuration (${\bf H} \parallel c$) is approached by the field-tilt, the AFM order in real space is found close to the FFLO nodal planes in contrast to the high field behavior in ${\bf H} \perp c$ case. Further, in the field v.s. temperature ($H$-$T$) phase diagram, the AFM order reduces, at a higher angle, to an AFM quantum critical point (QCP) lying at a {\it lower} field than $H_{c2}(0)$ as a consequence of competition between the field dependences of the nesting condition and of PPB. These features of the AFM order and the resulting $H$-$T$ phase diagram strikingly coincide with those seen in a recent NMR measurement on CeCoIn$_5$ in tilted field configurations. 
\end{abstract}


%\keyword{}

\maketitle

\section{I. Introduction}

Recently, the quasi two-dimensional heavy-fermion superconductor CeCoIn${\rm_{5}}$ has attracted much attention due to a possible realization of a Fulde-Ferrell-Larkin-Ovchinnikov (FFLO) superconducting (SC) state \cite{FF,LO} in its high-field low-temperature (HFLT) SC phase \cite{Bianch}. This new SC phase is separated through a second order transition on $H^*(T)$ from the familiar Abrikosov vortex lattice state and has been examined repeatedly in the field configuration ${\bf H} \perp c$ parallel to the SC planes \cite{Adachi}. The experimental fact in ${\bf H} \perp c$ that this new SC phase is extremely sensitive to both the magnetic \cite{Tokiwa1} and nonmagnetic \cite{Tokiwa2} impurity dopings implies \cite{R_Ikeda_imp} that, prior to the doping, this phase is spatially {\it inhomogeneous}. Further, an observed square-root ($\sim \sqrt{H-H^*}$) dependence of the internal field in a NMR measurement \cite{recent_NMR} has been consistent with the picture \cite{Adachi,GL_FFLO2} that the HFLT phase includes a FFLO spatial modulation {\it parallel} to the field. It should be kept in our mind that a similar HFLT phase also appears \cite{Bianch,GL_FFLO2,NMR06} in the perpendicular field configuration ${\bf H} \parallel c$ over a narrower field range. 

On the other hand, neutron scattering measurements in ${\bf H} \perp c$ have revealed the existence of an incommensurate AFM order within the HFLT SC phase \cite{Kenz1,Kenz2}
%, while any magnetic order has been absent in ${\bf H} \parallel c$
. The detected \cite{Kenz1,Kenz2} staggered moment ${\bf m}$ is parallel to the c-axis, and its incommensurate wavevector is parallel to [1,1,0] or [1,-1,0] irrespective of the ${\bf H}$-direction. This AFM ordering should be closely related to the AFM quantum critical behavior near the mean-field SC transition field $H_{c2}(0)$ observed not only in CeCoIn${\rm_{5}}$ in ${\bf H} \perp c$ and ${\bf H} \parallel c$ \cite{critical_Ce1,critical_Ce2,critical_Ce3} but also in pressured CeRhIn${\rm_{5}}$ \cite{critical_CeRh}, NpPd${\rm_{5}}$Al${\rm_{2}}$ \cite{critical_Np}, and Ce${\rm_{2}}$PdIn${\rm_{8}}$ \cite{critical_Ce2Pd}. 

It is noticeable that this high field AFM order does not appear outside the HFLT phase, because conventional theories in zero field suggest that the AFM order is suppressed by a nonvanishing value of the SC excitation gap \cite{AFM_gap1,AFM_gap2}. To explain why, in nonzero magnetic fields, the AFM order favors coexistence with the SC order, several pictures have been proposed so far \cite{Hatake,Hatakeyama,Yanase1,Machida,Ilya}. The common point of view to these theories is that the AFM order is enhanced by the $d_{x^{2}-y^{2}}$-wave \cite{dx2y2} pairing symmetry and a strong PPB effect. As will be discussed at the end of the present manuscript, on the other hand, there are crucial differences between those existing theories. 

In the present work, we focus on the intermediate field configurations connecting between the ${\bf H} \perp c$ and ${\bf H} \parallel c$ cases, motivated by several experiments performed in magnetic fields tilted from the basal ($a$-$b$) plane. Neutron scattering measurements \cite{nutron_kaiten} have discovered that the 17$^\circ$ rotation of the field away from the basal plane results in disappearance of the AFM order and have indicated that the staggered moment ${\bf m}$ remains fixed along the $c$-axis while the field is tilted. 
On the other hand, the magnetostriction experiments \cite{jiwai} and the magnetization measurements \cite{jika} have shown that the HFLT phase disappears at a larger angle, 20$^\circ$, which, by being combined with the neutron result \cite{nutron_kaiten}, suggests that the FFLO state with no AFM order is realized in a narrow range of the tilt angle. More recently, NMR data \cite{Kumagaionly} obtained by tilting the field direction from the $a$-$b$ plane have led to several nontrivial pictures on the HFLT phase. First of all, a separation of the AFM ordered region from the HFLT phase has been clearly seen even for the 7$^\circ$ rotation: The resulting AFM ordered region existing only within the HFLT phase is, in the $H$-$T$ phase diagram, narrower than the region of the HFLT one. Further, as the field direction is tilted, it is first lost from the {\it higher} fields and higher temperature side of the HFLT phase. This disappearance of the AFM order from higher fields suggests that an AFM quantum critical point (QCP) to be realized at a higher angle should lie at a lower field than $H_{c2}(0)$. This seems to be closely related to the experimental fact \cite{Kasahara,critical_Ce2} that the apparent AFM QCP in ${\bf H} \parallel c$ lies at a lower field than $H_{c2}(0)$. 
In addition, the NMR data in tilted fields \cite{Kumagaionly} suggest an AFM order lying, in the real space, in the vicinity of the FFLO nodal plane on which the SC order parameter vanishes in contrast to the picture seen in ${\bf H} \perp c$ that, at least in higher fields, the AFM order basically favors the spatial region with a nonvanishing SC order parameter \cite{recent_NMR,Hatake,Hatakeyama,aoyama}. 

In this paper, we develop a theory addressing possible HFLT phases of $d$-wave superconductors with strong PPB effects in the tilted field configurations by extending the treatments in Refs.\cite{Hatakeyama,aoyama}. To simplify theoretical analysis and make it easier to understand implication of the obtained results, two kinds of approaches for examining the angular dependences will be performed separately: one is based on deriving the Ginzburg-Landau(GL) mean field free energy, which takes a form of an expansion in the SC order parameter but fully includes both the paramagnetic and orbital pair-breaking effects, from an electronic Hamiltonian for an uniaxial Q2D model superconductor. There, effects of the orbital pair-breaking enhanced with the tilt of the field direction on the FFLO and AFM orderings are stressed. The other is the Pauli-limited model based on a tight-binding electronic Hamiltonian in which the resulting SC free energy fully includes the SC order parameter, while the orbital pair-breaking is neglected so that the SC order parameter is assumed to be homogeneous in the plane perpendicular to the field. 
%The latter approach becomes useful in studying the details of the resulting AFM% order in ${\bf k}$-space. 
It is found that the phase diagrams we obtain in the tilted field configurations become consistent with the experimentally observed one \cite{Kumagaionly}. 

This manuscript is organized as follows. In sec.II, we derive the GL mean field free energy by including both the paramagnetic and orbital pair-breaking effects together with the AFM order and primarily explain how the two (AFM and FFLO) orders induced by PPB are affected by the fild-tilt. In sec.III, the approach in the Pauli limit for the same issue is explained to discuss details of changes of the AFM order occurring when tilting the field. In summary, the obtained pictures on angular dependences of the HFLT phase of CeCoIn$_5$ are discussed, and our theory is compared with others \cite{Yanase1,Machida,Ilya} focusing on the parallel field case.

\section{II. Microscopic Ginzburg-Landau Approach}

In this section, the mean-field GL free energy for an uniaxial $d$-wave superconductor will be derived based on a Q2D microscopic Hamiltonian by incorporating both the paramagnetic and the orbital pair breaking effects and will be used to study how the resulting magnetic phase diagram, in particular the AFM order in the FFLO phase corresponding to the HFLT phase of CeCoIn$_5$, in our theory is affected by the tilt of the applied magnetic field from the basal plane. 
For simplicity of our analysis, the interaction terms will be treated from the outset in the mean field approximation. Then, our starting electronic Q2D Hamiltonian can be expressed, as given elsewhere \cite{aoyama}, in the form ${\cal H}={\cal H}_{\rm kin}+{\cal H}_{\rm SC}+{\cal H}_{\rm AF}$, where 
\begin{eqnarray}
{\cal H}_{\rm kin} &=& d\sum^{}_{\sigma ,j}\int_{}^{}d^{2} \mathbf{r}_{\perp}  \Biggl[  (\psi_{j}^{(\sigma)}(\mathbf{r}_{\perp}))^{\dag}%
\Bigl[ \, \xi(-i\nabla_\perp + e \mathbf{A}_\perp)-\sigma I \Bigr] \psi_{j}^{(\sigma)}(\mathbf{r}_{\perp}) \nonumber \\ 
&-& \frac{J}{2} \Bigl[(\psi_{j}^{(\sigma)}(\mathbf{r}_{\perp}))^{\dag} \psi_{j+1}^{(\sigma)}(\mathbf{r}_{\perp}) + {\rm H.c}\Bigr] \Biggr]% %
\label{J}, 
\label{q2dh}
\end{eqnarray}
with
\begin{eqnarray}
\psi_{j}^{(\sigma)}(\mathbf{r}_{\perp})=\frac{1}{\sqrt{V}}\sum^{}_{\mathbf{p}}\hat{c}_{\mathbf{p},\sigma}e^{i(\mathbf{p}_{\perp}\cdot \mathbf{r}_{\perp} + ip_{z}dj)}, 
\end{eqnarray}
and the mean field interaction terms on superconductivity ${\cal H}_{\rm SC}$ and antiferromagnetism ${\cal H}_{\rm AF}$ will be introduced below. 
The index $j$ is the label of the SC layers, $d$ is the interlayer distance in the $c$-direction, $\sigma(=\pm1)$ denotes the spin projection, $J$ represents the interlayer hopping integral, $\xi(\mathbf{p})$ is the kinetic energy measured from the Fermi energy $\mu$ in two-dimensional (2D) limit with $\mu > J$, $V$ is the system's volume, and $I=g(\theta)\mu_{\rm B}H$ is the Zeeman energy expressed with the Bohr magneton $\mu_{\rm B}$ and the angle-dependent $g$-factor $g(\theta)$. We shall introduce an uniaxial anisotropy and the resulting angle-dependence of the $g$-factor, because the real CeCoIn$_5$ shows such a remarkable anisotropy of the magnetic susceptibility \cite{taijiritu}. As a model, we assume the following form $g(\theta)=\sqrt{g_a^2 {\rm cos}^2\theta + g_c^2 {\rm sin}^2\theta}$, where $g_j$ is the $g$-factor for an applied field ${\bf H}$ in the $j$-direction, and $g_a=g_b$. The unit $\hbar=c=k_{\rm B}=1$ will be used throughout this paper.
The coordinates $\mathbf{r}=(x,y,z)$ will be often used which implies the coordinates $(\mathbf{r}_{\perp}, dj)$ in the $a-b-c$ crystal frame. That is, $x\mathchar`-$, $y\mathchar`-$ and $z\mathchar`-$axes are taken along the $a\mathchar`-$, $b\mathchar`-$ and $c\mathchar`-$axes, respectively.

To describe a superconductor in a magnetic field tilted away from the $a$-$b$ plane, we use a new rotated frame $(\tilde{x}\mathchar`-\tilde{y}\mathchar`-\tilde{z})$ defined by rotating the crystal frame $(x\mathchar`-y\mathchar`-z)$ around the $x\mathchar`-$axis. It is expressed as
\begin{eqnarray}
\tilde{x}=x, \,\,\,\, \tilde{y}=y\cos \theta + z\sin \theta, \,\,\,\, \tilde{z}=-y\sin \theta + z\cos \theta,
\label{eq:–¼'O} 
\end{eqnarray} 
where the magnetic field is $\mathbf{H}= H(\mathbf{\hat{{y}}}\cos \theta + \mathbf{\hat{{z}}}\sin \theta)=H\mathbf{\hat{{\tilde{y}}}} $. 
Then, in the type I\hspace{-.1em}I limit with no spatial variation of the flux density, the vector potential is represented simply by
\begin{eqnarray}
\mathbf{A}(\mathbf{r})=(H {\tilde z}, \,\,\, 0, \,\,\, 0)
\end{eqnarray}
in the $({\tilde x}, \,\,\,{\tilde y}, \,\,\, {\tilde z})$ frame. 

\begin{figure}[t]
\scalebox{0.6}[0.6]{\includegraphics{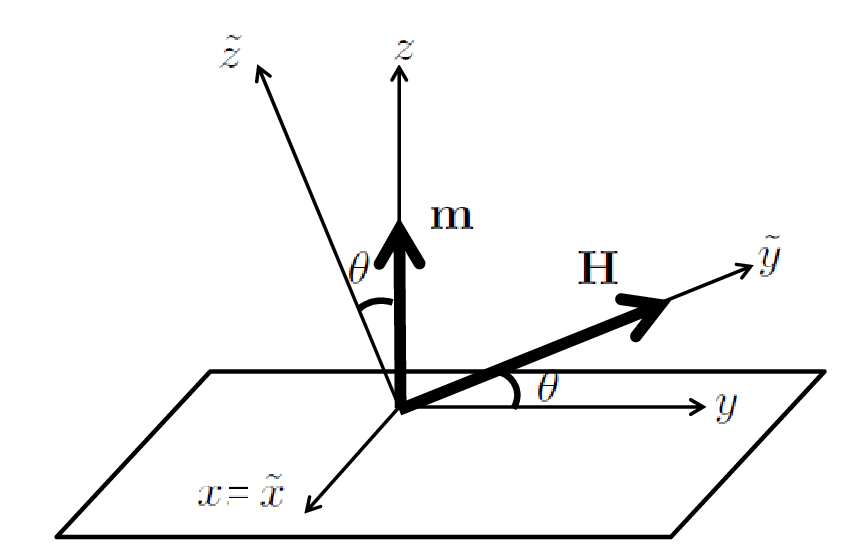}}
%\scalebox{0.3}[0.3]{\includegraphics{}}
\caption{Coordinate frames used in the present calculations. The $x\mathchar`-y\mathchar`-z$ frame corresponds to the crystal $a\mathchar`-$, $b\mathchar`-$ and $c\mathchar`-$ frame of an uniaxial crystal and the frame $\tilde{x}\mathchar`-\tilde{y}\mathchar`-\tilde{z}$ with the magnetic field ${\bf H}$ in the $\tilde{\bf y}$-direction is obtained by the $\theta$-rotation of the $x\mathchar`-y\mathchar`-z$ frame about the $x\mathchar`-$axis. According to Ref.\cite{nutron_kaiten} on CeCoIn$_5$, the orientation of the moment $\mathbf{m}$ is assumed to be locked in the $c$ direction irrespective of the ${\bf H}$-direction.}
\label{fig:1}
\end{figure}

The second term of eq.(\ref{q2dh}) represents an attractive interaction between quasiparticles, which leads to superconductivity, and, in the mean field approximation, may be expressed as
\begin{eqnarray}
{\cal H}_{\rm SC}=
\frac{1}{|g|} \sum^{}_{\mathbf{q}} |\Delta(\mathbf{q})|^{2} 
-\sum^{}_{\mathbf{q}}\left( \Delta(\mathbf{q}) \hat{\Psi}^{\dag}(\mathbf{q}) +{\rm H.c.} \right),
\end{eqnarray} 
with 
\begin{eqnarray}
\hat{\Psi}(\mathbf{q}) &=& \frac{1}{2} \sum^{}_{\mathbf{p},\alpha,\beta} (-i\hat{\sigma}_{y})_{\alpha,\beta} \, w_{\mathbf{p}} \, 
\hat{c}_{-\mathbf{p}+\frac{\mathbf{q}}{2},\alpha} \hat{c}_{\mathbf{p}+\frac{\mathbf{q}}{2},\beta}, \nonumber \\ 
\Delta(\mathbf{q}) &=& |g|\langle \hat{\Psi}(\mathbf{q}) \rangle.
\end{eqnarray}

Here, $\hat{\sigma}_{i}(i=x,y,z)$ are the Pauli matrices. The SC pairing symmetry is represented by the pairing function $w_{\mathbf{p}}$, and, in the case of $d_{x^{2}-y^{2}}$-pair, the identity $w_{\mathbf{p}+\mathbf{Q}_{0}}=-w_{\mathbf{p}}$ is satisfied, where $\mathbf{Q}_{0}=(\pi/a,\pi/a,\pi/d)$ is the commensurate nesting vector represented with the lattice constant $a$ in the $a$-$b$ plane. After this identity has been used in the analytic treatment, $w_{\mathbf{p}}$ will be replaced by its linearlized form $\sqrt{2}(\hat{p}_{x}^{2} - \hat{p}_{y}^{2})$ to perform the angle-average over the Fermi surface. 

The third term of eq.(\ref{q2dh}) is the AFM interaction term and, in the mean field approximation, takes the form 
\begin{eqnarray}
{\cal H}_{\rm AFM}=\frac{1}{U}\sum^{}_{\mathbf{q}}|\mathbf{m}(\mathbf{q})|^{2}
-\sum^{}_{\mathbf{q}}\left(\mathbf{m}(\mathbf{q}) \cdot \hat{\mathbf{S}}^{\dag}(\mathbf{q}) +{\rm H.c.} \right),
\end{eqnarray} 
with
\begin{eqnarray}
\hat{\mathbf{S}}(\mathbf{q})&=& \sum^{}_{\mathbf{p},\alpha,\beta} 
\hat{c}_{\mathbf{p},\alpha}^{\dag} \, (\mathbf{\hat{\boldsymbol{\sigma}}})_{\alpha,\beta} \, 
\hat{c}_{\mathbf{p}+\mathbf{Q}_{0}+\mathbf{q},\beta}, \nonumber \\ 
\mathbf{m}(\mathbf{q}) &=& U\langle \hat{\mathbf{S}}(\mathbf{q}) \rangle .
\end{eqnarray}
where the coupling constant $U$ is assumed to be positive. 
Within the present model, an incommensurate nesting property will be incorporated in the dispersion relation in the manner  
\begin{eqnarray}
\xi(\mathbf{p}+\mathbf{Q}_{0})=-\xi(\mathbf{p}) + \delta_{\rm IC} T_{c}, 
\label{AFM1}
\end{eqnarray}
where the deviation from a perfect nesting is represented by a constant parameter $\delta_{\rm IC}$. 
Then, we define the velocity $\mathbf{v_{p}}=d\xi(\mathbf{p}_\perp) /d \mathbf{p}_\perp + J {\rm sin}(p_z d) {\hat z}$ 
and we have the relation 
\begin{equation}
\mathbf{v}_{\mathbf{p}_\perp+\mathbf{Q}_{0}}=-\mathbf{v}_{\mathbf{p}_\perp}. 
\label{AFM2}
\end{equation}

We note that the gap function $\Delta(\mathbf{q})$ and
the staggered field $\mathbf{m}(\mathbf{q})$ play the roles of SC and AFM order parameters, respectively. Based on the previous works \cite{R_Ikeda_imp, Hatake,Hatakeyama,GL_FFLO1,GL_FFLO2}, we assume that the SC order parameter has a one-dimensional modulation of the Larkin-Ovchinnikov type \cite{LO} parallel to the applied field ${\bf H}$ 
\begin{eqnarray} 
\Delta(\mathbf{r})=|\Delta|\varphi_{0}(\tilde{z},\tilde{x})\sqrt{2} \cos( q_{\rm LO} \tilde{y} ),
\label{LOgap}
\end{eqnarray}
which corresponds to that in the HFLT phase of CeCoIn$_5$. Here, the FFLO wavenumber $q_{\rm LO}$ plays the role of the order parameter representing the presence of a FFLO modulation and vanishes with the square-root field dependence $\sim \sqrt{H-H^*(T)}$ (see sec.I) at the transition field $H^*$ to the ordinary Abrikosov lattice state with $q_{\rm LO}=0$. The ordinary Abrikosov vortex lattice is expressed by the function 
$\varphi_{0}$ belonging to the lowest Landau level. In the present tilted field configuration, it takes the form  
\begin{eqnarray} 
\varphi_{0}(\tilde{z},\tilde{x})=\sqrt{\frac{k}{\sqrt{\pi}}} \sum^{\infty }_{n=-\infty } 
\exp{\biggl[i \Bigl( \frac{nk\Gamma(\theta) }{r_{\rm H}} \tilde{x} + \frac{\pi n^{2}}{2} \Bigr)
-\frac{1}{2} \Bigl( \frac{1}{r_{\rm H} \Gamma(\theta)} \tilde{z} + nk \Bigr)^{2} \biggr]} 
\end{eqnarray}
with integer $n$, where $r_{\rm H}=1/\sqrt{2eH}$, and the angle-dependent factor $\Gamma(\theta)$ is associated with the material anisotropy and described as
\begin{eqnarray}
\hspace{-1em}
\Gamma^{4}(\theta) &=& \sin^{2}\theta + \frac{1}{\gamma^{2} } \cos^{2}\theta, \nonumber \\ 
\gamma &=& \sqrt{  \frac{ \left\langle v_{x}^{2}\right \rangle_{\rm FS}}{ \left\langle v_{z}^{2}\right \rangle_{\rm FS}} }
= \frac{2\sqrt{1-J/\mu}}{\pi J/\mu}.
\label{gamma}
\end{eqnarray}

As shown in our previous works \cite{Hatake,Hatakeyama}, the AFM order parameter should have a spatial modulation parallel to ${\bf H}$ through a coupling term $f^{(2,2)}_{\Delta,m}$ in the free energy with the SC order parameter. Then, we take 
\begin{eqnarray}
\mathbf{m}(\mathbf{r})= \sum^{}_{\mathbf{q}} |\mathbf{m}| e^{i\mathbf{q} \cdot \mathbf{r}} \sqrt{2}\cos( q_{\rm LO} \tilde{y} + \phi ).
\label{eq:afm_order}
\end{eqnarray}
for the AFM order parameter. For simplicity, we focus on the situation in which the FFLO modulation wavenumber $q_{\rm LO}$ is much smaller than that of the AFM modulation, so that {\it nonlocal} couplings between the AFM and FFLO orders may be negligible. Instead, through the last factor ${\rm cos}(q_{\rm LO} \tilde{y} + \phi)$ in eq.(\ref{eq:afm_order}), just the local coupling between the two orders stemming by the FFLO modulation parallel to the magnetic field will be taken into account \cite{Hatakeyama}. 

As illustrated in Fig.2, in the case of $\phi=0$, the AFM order primarily appears in the region where $|\Delta|$ is maximal, while it appears, when $\phi=\pi/2$, primarily in the vicinity of the FFLO nodal plane on which $\Delta=0$. Hereafter, we call the former as the {\it in-phase} configuration and the latter as the {\it out-of-phase} one. 
For the moment, we assume the direction $\mathbf{m}$ of the AFM moment to be locked along the $z\mathchar`-$axis corresponding to the $c$-axis of the Q2D material (see the caption of Fig.1 and sec.I). In this case, we have
$
\mathbf{m}(\mathbf{r})= 
m(\mathbf{r}) \mathbf{\hat{\boldsymbol{z}}}=
m(\mathbf{r}) (\mathbf{\hat{\boldsymbol{\tilde{y}}}} \, \, \sin \theta + \mathbf{\hat{\boldsymbol{\tilde{z}}}} \, \, \cos \theta) 
$.

\begin{figure}[t]
\scalebox{0.6}[0.6]{\includegraphics{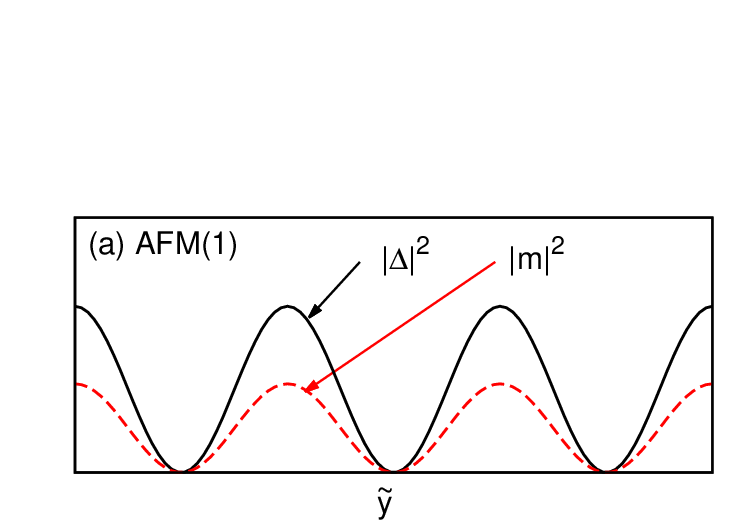}}
%\scalebox{0.3}[0.3]{\includegraphics{}}
\scalebox{0.6}[0.6]{\includegraphics{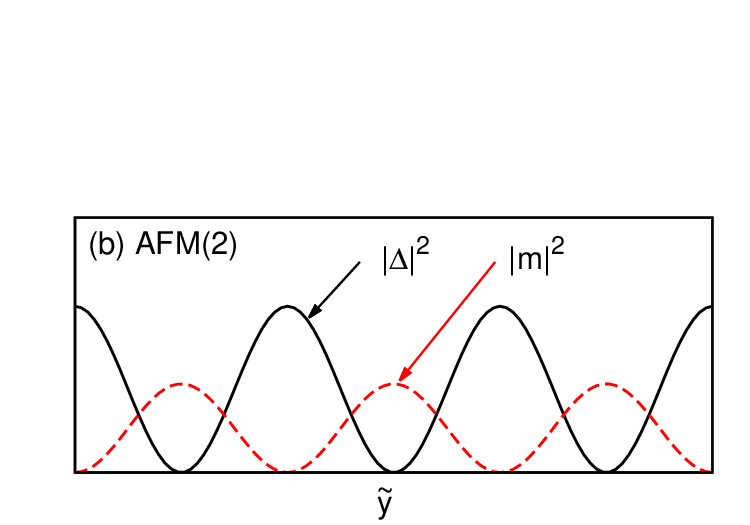}}
\caption{Typical configurations in real space of the AFM order (dashed (red) curve) in the FFLO state with one-dimensional spatial modulation of the amplitude $|\Delta|$ of the SC order parameter (solid (black) curve) parallel to the field. In the in-phase structure (a), the AFM order favors coexistence with the SC order, while, in the out-of-phase (b), it tends to lie around the nodal planes, on which $|\Delta|=0$, of the FFLO modulation (see the text). These states correspond to the $\phi=$ 0, and $\pi/2$ case in Eq.(\ref{eq:afm_order}), respectively.}
\label{fig:afm_structure}
\end{figure}

%%%% GL %%%%%%%%%%%%%
\subsection{\it Ginzburg-Landau free energy}

The mean-field free energy density for the Hamiltonian defined above 
is given by
\begin{eqnarray}
f_{\rm GL}(\Delta,m,q_{\rm LO}) = - V^{-1} T \ln {\rm Tr}_{c,c^{\dagger}}(\exp [ -( {\cal H}_{0}+{\cal H}_{\rm SC}+{\cal H}_{\rm AFM})/T]).
\end{eqnarray} 
In the present situation including the AFM and FFLO orders, we consider the following Ginzburg-Landau(GL) form of the free energy density expressed in powers of the order parameters $|\Delta|$ and $m \equiv |{\bf m}|$ 
\begin{eqnarray}
f_{\rm GL}(\Delta,m,q_{\rm LO}) &=& f^{(2)}_\Delta(q_{\rm LO}) + f^{(4)}_\Delta(q_{\rm LO}) + f^{(6)}_\Delta \nonumber \\
&+& f^{(2)}_m + f^{(4)}_m + f^{(2,2)}_{\Delta \, m}, 
\end{eqnarray} 
where
\begin{eqnarray}
f^{(2)}_\Delta(q_{\rm LO}) &=& f^{(2,0)}_\Delta + f^{(2,2)}_\Delta q_{\rm LO}^2 + f^{(2,4)}_\Delta q_{\rm LO}^4, \nonumber \\
f^{(4)}_\Delta(q_{\rm LO}) &=& f^{(4,0)}_\Delta + f^{(4,2)}_\Delta q_{\rm LO}^2 + f^{(4,4)}_\Delta q_{\rm LO}^4. 
\label{lo24} 
\end{eqnarray}
Since the high field $H_{c2}$-transition is discontinuous in the case with strong PPB \cite{Adachi}, we assume that the SC order is rigid enough and thus, is unaffected by the AFM ordering. That is, we determine the SC energy gap by focusing on the $m$-independent terms. In our GL approach taking account of the orbital pair-breaking, other higher order terms in $\Delta$ have been neglected \cite{Hatakeyama,aoyama}. We have repeatedly checked that, in low temperatures and high fields of our interest, the conditions $f^{(4)}_\Delta < 0$, and $f^{(6)}_\Delta > 0$ are always satisfied so that the $H_{c2}$-transition is discontinuous, while truncating the GL expansion in the six-th order in $|\Delta|$ is permitted. 
On the other hand, the FFLO transition line, i.e., the onset of the FFLO modulation of the SC order parameter is determined through appearance of a nonvanishing $q_{\rm LO}$ according to the expressions (\ref{lo24}). 

Here, we should mention that higher order terms in $q_{\rm LO}^2$ will be neglected, as in Ref.\cite{Hatakeyama}, in other free energy terms including the AFM order parameter $m$ such as the coupling term $f^{(2,2)}_{\Delta \, m}$ of the AFM and SC orders (see also the sentence below eq.(\ref{eq:afm_order})). Later, we will argue that this neglect of $q_{\rm LO}^2$-corrections, called as the {\it local approximation} in Ref.\cite{Hatakeyama}, does not change our main result on the resulting phase diagram. 

Further, it is found that the sign of $f^{(2,2)}_{\Delta \, m}$, proportional to $|\Delta|^2$, is negative for the tilt angles with the AFM order at finite temperatures. This sign favors the in-phase structure illustrated in Fig.2 (a) of the AFM order in real space. However, it will be shown later that this result may be an artifact of the use of the GL expansion in $\Delta$. 

%%%%%%%%%%%%%%%%%%%%%%%%%%%%%%%%%%%%%%%%%%
\subsection{\it GL coefficients}

Now, we turn to calculation of the coefficient of each term in the GL free energy density. 
To obtain each GL coefficient, we apply the semiclassical approximation 
\begin{eqnarray}
{\cal G}_{\varepsilon_{n},\sigma}^{(H)}(\mathbf{r},\mathbf{r}')\simeq {\cal G}_{\varepsilon_{n},\sigma}(\mathbf{r}-\mathbf{r}')
\exp{\left( ie\int_{\mathbf{r}}^{\mathbf{r}'}d\mathbf{s}\cdot \mathbf{A}(\mathbf{s}) \right) }
\end{eqnarray}
for the normal Green's function ${\cal G}_{\varepsilon_n,\sigma}^{(H)}$ in a magnetic field, where ${\cal G}_{\varepsilon_{n},\sigma}(\mathbf{r}-\mathbf{r}')$ appearing in r.h.s. is the Green's function in the case with no orbital pair breaking of the magnetic field, and its Fourier transformation is expressed by 
\begin{eqnarray}
{\cal G}_{\varepsilon_{n,\sigma}}(\mathbf{p})=[i\varepsilon_{n}-\xi(\mathbf{p}_\perp) + J (1-{\rm cos}(p_zd)) + \sigma I]^{-1}, 
\end{eqnarray}
where $\varepsilon_{n}=(2n+1)\pi T$ is the fermion's Matsubara frequency. Further, the orbital pair-breaking effect is incorporated by the gradient $\mathbf{\Pi}=-i\nabla+2e\mathbf{A}(\mathbf{r})$ operating on the pair-fields through the formula 
\begin{eqnarray}
\exp \biggl( 2ie\int_{\mathbf{r}}^{\mathbf{r}'}d\mathbf{s}\cdot \mathbf{A}(\mathbf{s}) \biggr) \Delta (\mathbf{r}')
= \exp \left(-i(\mathbf{r}-{\mathbf{r}'}) \cdot \mathbf{\Pi} \right) \Delta(\mathbf{r}).
\end{eqnarray}
Then, the quadratic, quartic and sixth-order terms in $\Delta$ of the GL free energy density are represented by

\begin{eqnarray}
f^{(2)}_\Delta (q_{\rm LO}) &=& \biggl< \Delta^{\ast}(\mathbf{r}) 
\biggl[ \frac{1}{|g|}-K^{(2)}_\Delta(\mathbf{\Pi}) \biggr] \Delta(\mathbf{r}) 
\biggr>_{\rm sp}, \nonumber \\
K^{(2)}_\Delta &=& \frac{T}{2} \sum^{}_{\varepsilon_{n}, \mathbf{p}, \sigma}  |w_{\mathbf{p}}|^{2} 
{\cal G}_{\varepsilon_{n},\sigma}(\mathbf{p}) {\cal G}_{-\varepsilon_{n}, -\sigma}(-\mathbf{p}+\mathbf{\Pi}), \nonumber \\
f^{(4)}_\Delta (q_{\rm LO}) &=& \Bigl< K^{(4)}_\Delta(\mathbf{\Pi}_{i})
\Delta^{\ast}(\mathbf{s}_{1})\Delta(\mathbf{s}_{2})\Delta^{\ast}(\mathbf{s}_{3})\Delta(\mathbf{s}_{4})
\Bigr|_{\mathbf{s}_{i}\rightarrow \mathbf{r}} 
\Bigr>_{\rm sp}, \nonumber \\
K^{(4)}_\Delta &=& \frac{T}{4} \sum^{}_{\varepsilon_{n},\mathbf{p}, \sigma}  |w_{\mathbf{p}}|^{4} 
{\cal G}_{\varepsilon_{n},\sigma}(\mathbf{p}) 
{\cal G}_{-\varepsilon_{n},-\sigma}(-\mathbf{p}+\mathbf{\Pi}^{\dag}_{1})
{\cal G}_{-\varepsilon_{n},-\sigma}(-\mathbf{p}+\mathbf{\Pi}_{2})
{\cal G}_{\varepsilon_{n},\sigma}(\mathbf{p}+\mathbf{\Pi}^{\dag}_{3}-\mathbf{\Pi}_{2}), \nonumber \\
f^{(6)}_\Delta (q_{\rm LO}) &=& \Bigl< K^{(6)}_\Delta(\mathbf{\Pi}_{i})
\Delta^{\ast}(\mathbf{s}_{1})\Delta(\mathbf{s}_{2})\Delta^{\ast}(\mathbf{s}_{3})\Delta(\mathbf{s}_{4})
\Delta^{\ast}(\mathbf{s}_{5})\Delta(\mathbf{s}_{6})
\Bigr|_{\mathbf{s}_{i}\rightarrow \mathbf{r}} 
\Bigr>_{\rm sp}, \nonumber \\
K^{(6)}_\Delta &=& \frac{T}{6} \sum^{}_{\varepsilon_{n}, \mathbf{p}, \sigma} |w_{\mathbf{p}}|^{6} 
{\cal G}_{\varepsilon_{n},\sigma}(\mathbf{p}) 
{\cal G}_{-\varepsilon_{n},-\sigma}(-\mathbf{p}+\mathbf{\Pi}^{\dag}_{1})
{\cal G}_{-\varepsilon_{n},-\sigma}(-\mathbf{p}+\mathbf{\Pi}_{6})
{\cal G}_{\varepsilon_{n},\sigma}(\mathbf{p}-\mathbf{\Pi}^{\dag}_{1}-\mathbf{\Pi}_{2}) \nonumber\\
&\times& {\cal G}_{-\varepsilon_{n},-\sigma}(-\mathbf{p}+\mathbf{\Pi}^{\dag}_{1}+\mathbf{\Pi}^{\dag}_{3}-\mathbf{\Pi}_{2})
{\cal G}_{\varepsilon_{n},\sigma}(\mathbf{p}-\mathbf{\Pi}_{6}+\mathbf{\Pi}^{\dag}_{5}).
\label{eq:sc_part} 
\end{eqnarray}
The concrete expressions of these terms are represented in Appendix.
In obtaining them, we need to rewrite the expression $\exp{(i A \mathbf{v}_{\mathbf{p}} \cdot \mathbf{\Pi}) }\Delta(\mathbf{r})$. To perform this, it will be represented in the rotated frame as follows:  
\begin{eqnarray}
\exp{(i A \mathbf{v}_{\mathbf{p}} \cdot \mathbf{\Pi}) }
&=& \exp{(i A \mathbf{\boldsymbol{\tilde{v}}}_{\mathbf{\boldsymbol{\tilde{p}}}}  \cdot \mathbf{\boldsymbol{\tilde{\Pi}}} )} \nonumber\\
&=& \exp{(i A \tilde{v}_{\tilde{p},\tilde{y}}  \tilde{\Pi}_{\tilde{y}} )}
\exp{(i A \mathbf{\boldsymbol{\tilde{v}}}_{\mathbf{\boldsymbol{\tilde{p}}},\perp}  \cdot \mathbf{\boldsymbol{\tilde{\Pi}}_{\perp}} )}
%\exp(i\rho V_{\mathbf{p},Y}\Pi_{Y}') \exp(i\rho \mathbf{V}_{\mathbf{p}}_{\perp }\cdot\mathbf{\Pi}'_{\perp}),
\label{eq:yzx}
\end{eqnarray}
where 
$\mathbf{\boldsymbol{\tilde{v}}}_{\mathbf{\boldsymbol{\tilde{p}}},\perp}=(\tilde{v}_{\tilde{p},\tilde{z}}  ,\tilde{v}_{\tilde{p},\tilde{x}} )$ and $\mathbf{\boldsymbol{\tilde{\Pi}}_{\perp}}=(\tilde{\Pi}_{\tilde{z}},\tilde{\Pi}_{\tilde{x}})$ represent the components perpendicular to the field ($\parallel {\tilde y}$-axis) of $\mathbf{v}_{\mathbf{p}}$ and $\mathbf{\Pi}$ in the rotated frame (see Fig.1). By introducing the creation and annihilation operators on the Landau levels representing possible vortex states 
\begin{eqnarray}
\tilde{\Pi}_{\pm}=\frac{r_{\rm H}}{\sqrt{2}}(\Gamma(\theta) \tilde{\Pi}_{\tilde{z}} \pm i \Gamma^{-1}(\theta) \tilde{\Pi}_{\tilde{x}}), 
\end{eqnarray}
we find
\begin{eqnarray}
\exp{(i A \mathbf{\boldsymbol{\tilde{v}}}_{\mathbf{\boldsymbol{\tilde{p}}},\perp}  \cdot \mathbf{\boldsymbol{\tilde{\Pi}}_{\perp}} )}
=\exp{\Bigl( -\frac{1}{2} |\mu|^{2} A^{2} \Bigr)} \exp(i\mu \tilde{\Pi}_{+} A)  \exp(i\mu^{\ast} \tilde{\Pi}_{-} A), 
\label{kidou}
\end{eqnarray}
where 
\begin{eqnarray}
\mu=\frac{\Gamma^{-1}(\theta) \tilde{v}_{\tilde{p},\tilde{z}} -i \Gamma(\theta) \tilde{v}_{\tilde{p},\tilde{x}}}{\sqrt{2} r_{\rm H} T_{c} }.
\label{eq:q}
\end{eqnarray}

Next, we calculate the GL terms associated with the AFM order parameter ${\bf m}$. The expression of the term quadratic in ${\bf m}$ is given by
\begin{eqnarray}
f^{(2)}_m = \biggl< 
\biggl[ 
\frac{1}{U}+ \sum^{2}_{j=1}  K^{(2)}_{m, \, j}(\mathbf{q}) 
\biggr] 
|m(\mathbf{r})|^{2}
\biggr>_{\rm sp} ,
\end{eqnarray} 
where ${\bf q}$ is the incommensurate part of the AFM wavevector which should be determined by minimizing the free energy, and the concrete expression of $K^{(2)}_{m, \, j}$ is 
\begin{eqnarray}
K^{(2)}_{m, \, j}(\mathbf{q}) &=&
\frac{A_{j} T}{2}\sum^{}_{\varepsilon_{n}, \mathbf{p}, \sigma} 
{\cal G}_{\varepsilon_{n},\sigma}(\mathbf{p}) {\cal G}_{\varepsilon_{n},\alpha_{j}}(\mathbf{p}+\mathbf{Q}_{0}+\mathbf{q}) \nonumber\\
&=& - A_{j} \pi TN(0) \sum^{}_{\varepsilon_{n}, \sigma}  \biggl\langle
\frac{is_{\epsilon}}{2i\varepsilon_{n} +(\sigma + \alpha_{j})I  - \delta_{\rm IC} T_{c}+  \mathbf{v}_{\mathbf{p}} \cdot \mathbf{q} }
\biggr\rangle_{\rm FS} \nonumber\\
&=& - A_{j} N(0) \int_{0}^{\infty}d\rho f(\rho, B_{j}) \Bigl\langle \cos{ \Bigl( \Bigl( - \delta_{\rm IC} 
+ \frac{\mathbf{v}_{\mathbf{p}} \cdot \mathbf{q}}{T_{c}}  \Bigr) \rho \Bigr) }\Bigr\rangle_{\rm FS}.
\label{eq:GL_m2_1}
\end{eqnarray}
Here, the angle brackets denote the Fermi surface average, $N(0)$ is the density of states at the Fermi energy, $s_{\epsilon}$ is the sign of $\varepsilon_{n}$, 
\begin{eqnarray}
f(x, y) = \frac{2\pi t}{\sinh{(2\pi tx)}} \cos{ \biggl( 2\frac{I}{T_{c}} y \biggr) },
\end{eqnarray}
with $t=T/T_{c}$, and the coefficients $\alpha_{j},A_{j},B_{j}$ are represented in Table \ref{tb:table1}.
Here, the identity
\begin{eqnarray}
\frac{1}{\alpha} = \int_{0}^{\infty}d\rho \exp{(-\alpha\rho)} & (\rm Re\; \alpha>0),
\label{eq:para}
\end{eqnarray}
was used in obtaining Eq. (\ref{eq:GL_m2_1}).

\begin{table}[htb]
%\begin{center}
\caption{Coefficients $\alpha_{j}$, $A_{j}$, $B_{j}$ in Eq. (\ref{eq:GL_m2_1}) and (\ref{eq:GL_m2_2}). }
  \begin{tabular}{cccc} \hline
    $j$ & $\alpha_{j}$ & $A_{j}$ & $B_{j}$ \\ \hline
    $1$ & $-\sigma$ & $\cos^{2}\theta$ & $0$ \\
    $2$ & $\sigma$ & $\sin^{2}\theta$ & $\rho$ \\ \hline
  \end{tabular}
  \label{tb:table1}
%  \end{center}
\end{table}
By using Eq. (\ref{eq:para}), the coupling constant $U$ is represented by
\begin{eqnarray}
\frac{1}{U} &=& N(0) \biggl( \ln{\frac{T}{T_{\rm N}} } + 2\pi T\sum_{\varepsilon_{n}>0} \frac{1}{\varepsilon_{n}} \biggr) \nonumber\\
&=& N(0)\biggl (\ln{ \frac{T}{T_{\rm N}} } + \int_{0}^{\infty}d\rho f(\rho, 0)  \biggr)
\end{eqnarray}
where $T_{\rm N}$ is the AFM transition temperature in the normal state. Then, the quadratic term in ${\bf m}$ is expressed by
\begin{eqnarray}
f^{(2)}_m=
N(0) T_{c}^{2}  \biggl[ \ln{ \frac{T}{T_{\rm N}} } + \int_{0}^{\infty}d\rho \biggl( f(\rho, 0) 
- \sum^{2}_{j=1} A_{j} f(\rho, B_{j}) \Bigl\langle \cos{ \Bigl( \Bigl( - \delta_{\rm IC} 
+ \frac{\mathbf{v}_{\mathbf{p}} \cdot \mathbf{q}}{T_{c}}  \Bigr) \rho \Bigr) }\Bigr\rangle_{\rm FS} \biggr) \biggr]
\biggl( \frac{|m|}{T_{c}} \biggr)^{2} .
\label{eq:GL_m2_2}
\end{eqnarray} 

Further, the term giving the coupling between the SC and magnetic orders is expressed by
\begin{eqnarray}
f^{(2,2)}_{\Delta \, m} =
\biggl< 
\biggl[
\sum^{4}_{j=1} K^{(2,2)}_{\Delta \, m, \, j}(\mathbf{\Pi}_{s},\mathbf{q}) 
\biggr]  \Delta^{\ast}(\mathbf{r})  \Delta(\mathbf{s}) 
\Bigr|_{\mathbf{s}\rightarrow \mathbf{r}} |m(\mathbf{r})|^{2}
\biggr>_{\rm sp},
\end{eqnarray} 
where the kernels $K^{(2,2)}$ with $j=1,2$ take the form 
\begin{eqnarray}
K^{(2,2)}_{\Delta \, m, \, j}(\mathbf{\Pi}_{s},\mathbf{q})
&=& \frac{A'_{j} T}{2} \sum^{}_{\varepsilon_{n}, \mathbf{p}, \sigma} |w_{\mathbf{p}}|^{2}
{\cal G}_{\varepsilon_{n},\sigma}(\mathbf{p}) 
{\cal G}_{\varepsilon_{n},\alpha'_{j}}(\mathbf{p}-\mathbf{Q}_{0}-\mathbf{q}) {\cal G}_{\varepsilon_{n},\sigma}(\mathbf{p}) 
{\cal G}_{-\varepsilon_{n},-\sigma}(-\mathbf{p}+\mathbf{\Pi}_{\mathbf{s}}), \nonumber\\
&=& A'_{j} \pi T N(0)  \sum^{}_{\varepsilon_{n}>0,\sigma, s_{\epsilon}}  
\biggl\langle |w_{\mathbf{p}}|^{2} 
\frac{is_{\epsilon}}{ (2i\varepsilon_{n} +(\sigma + \alpha'_{j})I  - \delta_{\rm IC} T_{c} - \mathbf{v}_{\mathbf{p}} \cdot \mathbf{q}) (2i\varepsilon_{n} +2\sigma I - \mathbf{v}_{\mathbf{p}} \cdot \mathbf{\Pi}_{\mathbf{s}})}  \nonumber\\
&\times & \biggl(
\frac{1}{2i\varepsilon_{n} +(\sigma + \alpha'_{j})I  - \delta_{\rm IC} T_{c} - \mathbf{v}_{\mathbf{p}} \cdot \mathbf{q}}
-\frac{1}{2i\varepsilon_{n} +2\sigma I - \mathbf{v}_{\mathbf{p}} \cdot \mathbf{\Pi}_{\mathbf{s}}}
\biggr) 
\biggr\rangle_{\rm FS} , \nonumber\\
&=& \frac{A'_{j} N(0)}{2} \sum^{}_{s_{\epsilon}} \int_{0}^{\infty} \prod^{3}_{i=1}d\rho_{i}
\biggl[
f\biggl(\sum^{3}_{i=1} \rho_{i}, B'_{j} \biggr)
\Bigl\langle |w_{\mathbf{p}}|^{2} 
\exp{ \Bigl( -is_{\epsilon} \Bigl( - \delta_{\rm IC} 
+ \frac{\mathbf{v}_{\mathbf{p}} \cdot \mathbf{q}}{T_{c}}  \Bigr) C'_{j} \Bigr) }
\exp{ \Bigl( -is_{\epsilon} \mathbf{v}_{\mathbf{p}} \cdot \mathbf{\Pi}_{\mathbf{s}} D'_{j} \Bigr) }\Bigr\rangle_{\rm FS} \nonumber\\
&+& f\biggl(\sum^{3}_{i=1} \rho_{i}, E'_{j} \biggr)
\Bigl\langle |w_{\mathbf{p}}|^{2} 
\exp{ \Bigl( -is_{\epsilon} \Bigl( - \delta_{\rm IC} 
+ \frac{\mathbf{v}_{\mathbf{p}} \cdot \mathbf{q}}{T_{c}}  \Bigr) F'_{j} \Bigr) }
\exp{ \Bigl( -is_{\epsilon}  \frac{\mathbf{v}_{\mathbf{p}} \cdot \mathbf{\Pi}_{\mathbf{s}}}{T_{c}}
 G'_{j} \Bigr) }\Bigr\rangle_{\rm FS}
\biggr],
\label{eq:GL_m2d2_1}
\end{eqnarray}
while, for $j=3,4$, this kernels are expressed, in terms of the property $w_{\mathbf{p}+\mathbf{Q}_{0}}=-w_{\mathbf{p}}$ on the $d_{x^2-y^2}$-wave pairing function, by
\begin{eqnarray}
K^{(2,2)}_{\Delta \, m, \, j}(\mathbf{\Pi}_{s},\mathbf{q}) &=&
- \frac{A'_{j} T}{2} \sum^{}_{\varepsilon_{n}, \mathbf{p}, \sigma} w_{\mathbf{p}} w_{\mathbf{p}+\mathbf{Q}_{0}}
{\cal G}_{\varepsilon_{n},\sigma}(\mathbf{p}) 
{\cal G}_{\varepsilon_{n},\alpha'_{j}}(\mathbf{p}-\mathbf{Q}_{0}-\mathbf{q}) 
{\cal G}_{-\varepsilon_{n},\alpha'_{j}}(-\mathbf{p}+\mathbf{Q}_{0}+\mathbf{q}+\mathbf{\Pi}_{\mathbf{s}}) \nonumber\\
&\times& {\cal G}_{-\varepsilon_{n},-\sigma}(-\mathbf{p}+\mathbf{\Pi}_{\mathbf{s}})  \nonumber\\
&=& - A'_{j} \pi T N(0)  \sum^{}_{\varepsilon_{n}>0,\sigma, s_{\epsilon}}  
\biggl\langle |w_{\mathbf{p}}|^{2} 
\frac{is_{\epsilon}}{ (2i\varepsilon_{n} +(\sigma + \alpha'_{j})I  - \delta_{\rm IC} T_{c} - \mathbf{v}_{\mathbf{p}} \cdot \mathbf{q}) } \nonumber\\
&\times& 
\frac{1}{(2i\varepsilon_{n} +(\sigma + \alpha'_{j}) I + \delta_{\rm IC} T_{c}+  \mathbf{v}_{\mathbf{p}} \cdot \mathbf{q})}
\biggl(
\frac{1}{2i\varepsilon_{n} +2\sigma I - \mathbf{v}_{\mathbf{p}} \cdot \mathbf{\Pi}_{\mathbf{s}} }
-\frac{1}{2i\varepsilon_{n} + 2\alpha'_{j} I + \mathbf{v}_{\mathbf{p}} \cdot \mathbf{\Pi}_{\mathbf{s}}}
\biggr) 
\biggr\rangle_{\rm FS} \nonumber\\
&=& \frac{A'_{j} N(0)}{2} \sum^{}_{s_{\epsilon}} \int_{0}^{\infty} \prod^{3}_{i=1}d\rho_{i}
\biggl[
f\biggl(\sum^{3}_{i=1} \rho_{i}, B'_{j} \biggr)
\Bigl\langle |w_{\mathbf{p}}|^{2} 
\exp{ \Bigl( -is_{\epsilon} \Bigl( - \delta_{\rm IC} + \frac{\mathbf{v}_{\mathbf{p}} \cdot \mathbf{q}}{T_{c}}  \Bigr) C'_{j} \Bigr) }
\exp{ \Bigl( -is_{\epsilon}  \frac{\mathbf{v}_{\mathbf{p}} \cdot \mathbf{\Pi}_{\mathbf{s}}}{T_{c}} D'_{j} \Bigr) }\Bigr\rangle_{\rm FS} \nonumber \\
&+& f\biggl(\sum^{3}_{i=1} \rho_{i}, E'_{j} \biggr)
\Bigl\langle |w_{\mathbf{p}}|^{2} 
\exp{ \Bigl( -is_{\epsilon} \Bigl( - \delta_{\rm IC} + \frac{\mathbf{v}_{\mathbf{p}} \cdot \mathbf{q}}{T_{c}}  \Bigr) F'_{j} \Bigr) }
\exp{ \Bigl( -is_{\epsilon}  \frac{\mathbf{v}_{\mathbf{p}} \cdot \mathbf{\Pi}_{\mathbf{s}}}{T_{c}} G'_{j} \Bigr) }\Bigr\rangle_{\rm FS}
\biggr].
\label{eq:GL_m2d2_2}
\end{eqnarray}
The coefficients $\alpha'_{j}$, $A'_{j}$, $B'_{j}$, $C'_{j}$, $D'_{j}$, $E'_{j}$, $F'_{j}$, $G'_{j}$ are represented in Table \ref{tb:table2}.

By using the above-mentioned mathematical tools, we obtain
\begin{eqnarray}
f^{(2,2)}_{\Delta \, m} &=&
\frac{3N(0)T_{c}^{2}}{2}  \sum^{4}_{j=1} A'_{j}   \int_{0}^{\infty} \prod^{3}_{i=1}d\rho_{i} 
\biggl[ f\biggl(\sum^{3}_{i=1} \rho_{i}, B'_{j} \biggr)
\Bigl\langle |w_{\mathbf{p}}|^{2} 
\cos{ \Bigl( \Bigl( -\delta_{\rm IC} + \frac{\mathbf{v}_{\mathbf{p}} \cdot \mathbf{q}}{T_{c}}  \Bigr) C'_{j} \Bigr) }
\exp{ \Bigl( - \frac{|\mu|^{2}}{2} D'_{j} \Bigr) }\Bigr\rangle_{\rm FS} \nonumber\\
&+& f\biggl(\sum^{3}_{i=1} \rho_{i}, E'_{j} \biggr)
\Bigl\langle |w_{\mathbf{p}}|^{2} 
\cos{ \Bigl( \Bigl( -\delta_{\rm IC} + \frac{\mathbf{v}_{\mathbf{p}} \cdot \mathbf{q}}{T_{c}}  \Bigr) F'_{j} \Bigr) }
\exp{ \Bigl( - \frac{|\mu|^{2}}{2} G'_{j} \Bigr) }\Bigr\rangle_{\rm FS}
\biggr]
\biggl( \frac{|\Delta|}{T_{c}} \biggr)^{2} 
\biggl( \frac{|m|}{T_{c}} \biggr)^{2} . 
\label{eq:GL_m2d2_3}
\end{eqnarray}

\begin{table}[htb]
%\begin{center}
\caption{Coefficients $\alpha'_{j}$, $A'_{j}$, $B'_{j}$, $C'_{j}$, $D'_{j}$, $E'_{j}$, $F'_{j}$, $G'_{j}$ in Eq. (\ref{eq:GL_m2d2_1}), (\ref{eq:GL_m2d2_2}) and (\ref{eq:GL_m2d2_3})}
  \begin{tabular}{ccccccccc} \hline
    $j$ & $\alpha'_{j}$ & $A'_{j}$ & $B'_{j}$ & $C'_{j}$ & $D'_{j}$ & $E'_{j}$ & $F'_{j}$ & $G'_{j}$\\ \hline
    $1$ & $-\sigma$ & $2\cos^{2}\theta$ & $\rho_{2}$ & $\rho_{1}+\rho_{3}$ & $\rho_{2}$ & $\rho_{2}+\rho_{3}$ & $\rho_{1}$ & $\rho_{2}+\rho_{3}$ \\
    $2$ & $\sigma$ & $2\sin^{2}\theta$ & $\rho_{2}$ & $\rho_{1}+\rho_{3}$ & $\rho_{2}$ & $\rho_{2}+\rho_{3}$ & $\rho_{1}$ & $\rho_{2}+\rho_{3}$ \\
    $3$ & $-\sigma$ & $\cos^{2}\theta$ & $\rho_{3}$ & $-\rho_{1}+\rho_{2}$ & $\rho_{3}$ & $\rho_{3}$ & $-\rho_{1}+\rho_{2}$ & $-\rho_{3}$ \\
    $4$ & $\sigma$ & $\sin^{2}\theta$ & $\sum^{3}_{i=1} \rho_{i}$ & $-\rho_{1}+\rho_{2}$ & $\rho_{3}$ & $\sum^{3}_{i=1} \rho_{i}$ & $-\rho_{1}+\rho_{2}$ & $-\rho_{3}$ \\ \hline
  \end{tabular}
  \label{tb:table2}
%  \end{center}
\end{table}

Finally, the quartic term $f^{(4)}_m$ in ${\bf m}$ is expressed as
\begin{eqnarray}
f^{(4)}_{m} = \biggl<
\biggl[ 
\sum^{5}_{j=1}  K^{(4)}_{m, \, j}(\mathbf{q})
\biggr] 
|m(\mathbf{r})|^{4}
\biggr>_{\rm sp},
\end{eqnarray} 
where 
\begin{eqnarray}
\hspace{-4em}
K^{(4)}_{m, \, j}(\mathbf{q}) &=&
\frac{A''_{j} T}{2} \sum^{}_{\varepsilon_{n}, \mathbf{p}, \sigma} 
{\cal G}_{\varepsilon_{n},\sigma}(\mathbf{p}) 
{\cal G}_{\varepsilon_{n},\alpha''_{j}}(\mathbf{p}+\mathbf{Q}_{0}+\mathbf{q})
{\cal G}_{\varepsilon_{n},\beta''_{j}}(\mathbf{p})
{\cal G}_{\varepsilon_{n},\gamma''_{j}}(\mathbf{p}+\mathbf{Q}_{0}+\mathbf{q})   \nonumber\\
&=&  N(0) A''_{j}
\int_{0}^{\infty} \prod^{3}_{i=1}d\rho_{i}
\biggl[ f\biggl(\sum^{3}_{i=1} \rho_{i}, B''_{j} \biggr) 
+f\biggl(\sum^{3}_{i=1} \rho_{i}, C''_{j} \biggr) 
\biggr]
\Bigl\langle 
\cos{ \biggl( \Bigl( -\delta_{\rm IC} + \frac{\mathbf{v}_{\mathbf{p}} \cdot \mathbf{q}}{T_{c}}  \Bigr) \biggl(\sum^{3}_{i=1} \rho_{i} \biggr) \biggr) }
\Bigr\rangle_{\rm FS}.
\label{eq:GL_m4}
\end{eqnarray}
Then, the corresponding term in the free energy is expressed as
\begin{eqnarray}
\hspace{-3em}
f^{(4)}_{m} &=& \frac{3 N(0) T_{c}^{2} }{2} 
\sum^{5}_{j=1} A''_{j}
\int_{0}^{\infty} \prod^{3}_{i=1}d\rho_{i}
\biggl[ f\biggl(\sum^{3}_{i=1} \rho_{i}, B''_{j} \biggr) 
+f\biggl(\sum^{3}_{i=1} \rho_{i}, C''_{j} \biggr) 
\biggr] \nonumber\\
&\times& \Bigl\langle 
\cos{ \biggl( \Bigl( -\delta_{\rm IC} + \frac{\mathbf{v}_{\mathbf{p}} \cdot \mathbf{q}}{T_{c}}  \Bigr) \biggl(\sum^{3}_{i=1} \rho_{i} \biggr) \biggr) }
\Bigr\rangle_{\rm FS}
\biggl( \frac{|m|}{T_{c}} \biggr)^{4}.
\label{eq:GL_m4_2}
\end{eqnarray}

\begin{table}[htb]
%\begin{center}
\caption{Coefficients $\alpha''_{j}$, $ \beta''_{j}$, $\gamma''_{j}$, $A''_{j}$, $B''_{j}$, $C''_{j}$ in Eq. (\ref{eq:GL_m4}) and (\ref{eq:GL_m4_2}) }
  \begin{tabular}{ccccccc} \hline
    $j$ & $\alpha''_{j}$ & $\beta''_{j}$ & $\gamma''_{j}$ & $A''_{j}$ & $B''_{j}$ & $C''_{j}$ \\ \hline
    $1$ & $-\sigma$ & $\sigma$ & $-\sigma$ & $\cos^{4}\theta$ & 0 & 0 \\
    $2$ & $\sigma$ & $\sigma$ & $\sigma$ & $\sin^{4}\theta$ & $\sum^{3}_{i=1} \rho_{i}$ & $\sum^{3}_{i=1} \rho_{i}$ \\
    $3$ & $\sigma$ & $\sigma$ & $-\sigma$ & $2\cos^{2}\theta \sin^{2}\theta$ & $\rho_{1}$ & $\rho_{1}+\rho_{3}$  \\
    $4$ & $\sigma$ & $-\sigma$ & $\sigma$ & $2\cos^{2}\theta \sin^{2}\theta$ & $\rho_{1}+\rho_{3}$ & $\rho_{1}$ \\
    $5$ & $\sigma$ & $-\sigma$ & $-\sigma$ & $-2\cos^{2}\theta \sin^{2}\theta$ & $\rho_{1}-\rho_{2}$ & $\rho_{1}-\rho_{2}$ \\ \hline
\end{tabular}
\label{tb:table3}
%\end{center}
\end{table}

The coefficients $\alpha''_{j}$, $ \beta''_{j}$, $\gamma''_{j}$, $A''_{j}$, $B''_{j}$, $C''_{j}$ are listed in Table \label{tb:table3}.

The resulting numerical calculation results are characterized by the Maki parameter 
\begin{eqnarray}
\alpha_{\rm M}(\theta) = \frac{\sqrt{2} H_{\rm orb}^\theta(0)}{H_{\rm P}^\theta(0)}
\end{eqnarray}
generalized to the case with the tilt angle $\theta$, which measures the relative strength of the paramagnetic and orbital pair-breaking effects at the angle $\theta$. Here, $H_{\rm P}^\theta(0)$ is the Pauli-limiting field at $T=0$ and at the angle $\theta$ and is defined as $\pi T_c/(2 e^{\gamma_{\rm E}} \mu_{\rm B} g(\theta))$, where $e^{\gamma_{\rm E}} = 1.77$ is the Euler constant, while the $T=0$ orbital-limiting field $H_{\rm orb}^\theta(0)$ satisfies 
$H_{\rm orb}^{\theta}(0) = H_{\rm orb}^{(\theta=0)}(0)/\gamma \Gamma^2(\theta)$. Both of these two limiting fields decrease with tilting the field direction from the $a$-$b$ plane. In our numerical calculations, we use the parameter values $g_c/g_a=2.1$, $\gamma=2.8$, and $\alpha_{\rm M}(0)=6.11$. 

%%%%%%%%%%%%%%%%%%%%%%%%%%%%%%%%%%
\begin{figure}[t]
\scalebox{0.6}[0.6]{\includegraphics{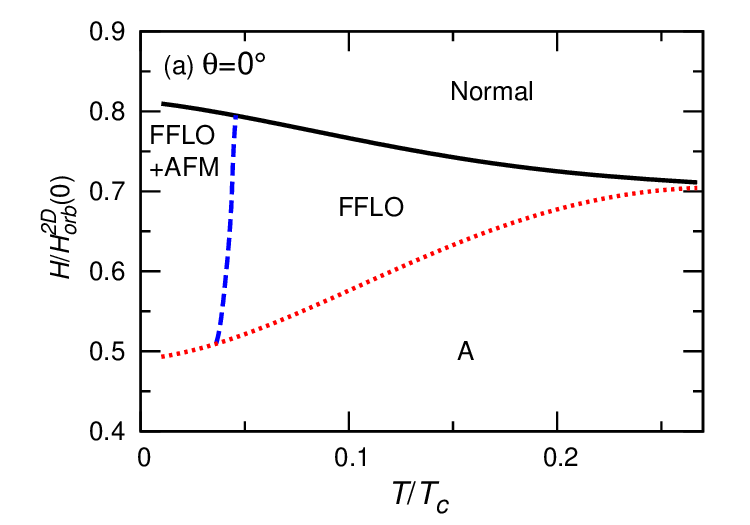}}
%\scalebox{0.3}[0.3]{\includegraphics{}}
\scalebox{0.6}[0.6]{\includegraphics{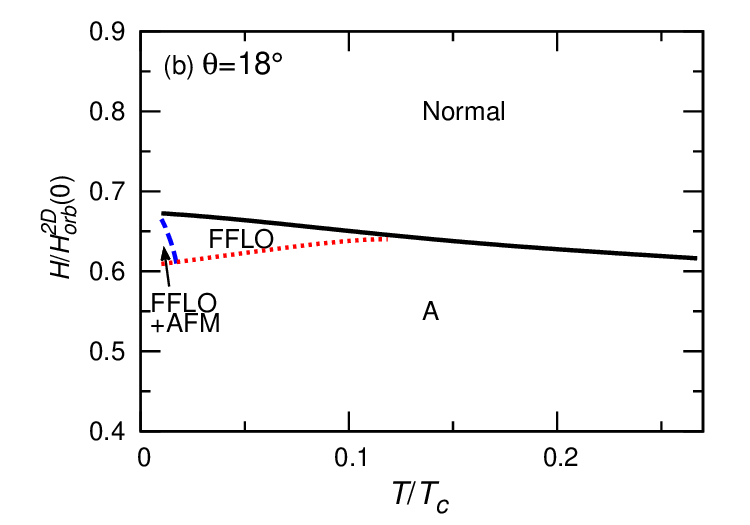}}
\scalebox{0.6}[0.6]{\includegraphics{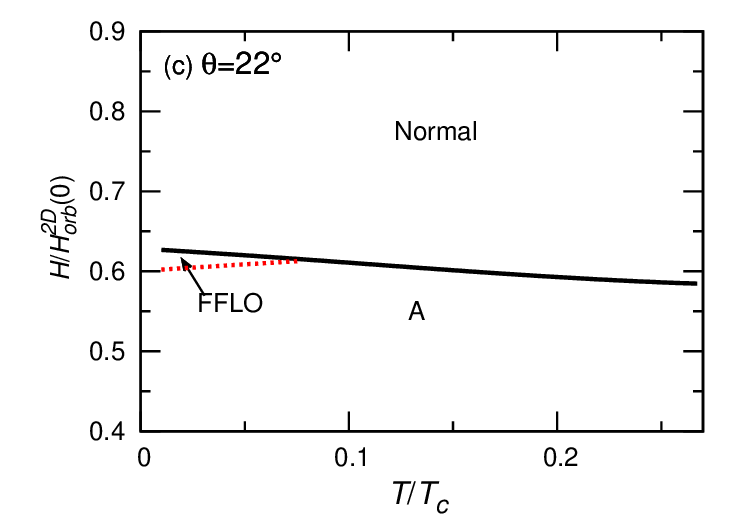}}
%\scalebox{0.3}[0.3]{\includegraphics{}}
\scalebox{0.6}[0.6]{\includegraphics{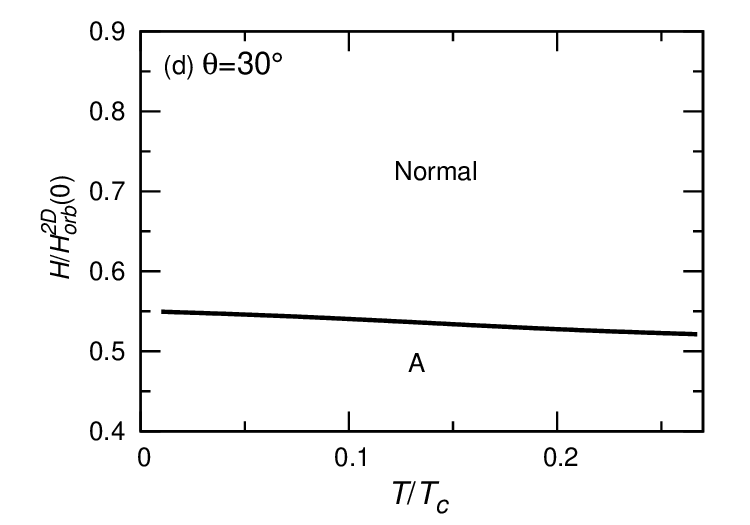}}
\caption{Angular dependence of the $H\mathchar`-T$ phase diagrams for $\theta= 0, 18, 22$ and $30$ case. The dotted (red) and dashed (blue) lines represent the transitions from the Abrikosov lattice (A) phase to the FFLO one and from the nonmagnetic FFLO state to the AFM-FFLO order, respectively, and both of them are second order transition lines, while the thick solid (black) one, which shows the $H_{c2}$ transition, is a discontinuous transition curve. In this calculation, the parameters $\alpha_{\rm M}(0) =6.11$, $\delta=0.001$, $\gamma=2.8$, $g_c/g_a=2.1$, and $T_{\rm N}/T_{\rm c}=0.0012$ are used.
}
\label{fig:gl}
\end{figure}

\subsection{\it Results}

Figure \ref{fig:gl} shows possible $H$ v.s. $T$ phase diagrams for tilt angles $\theta= 0, 18, 22$, and $30$ degrees, respectively. The dashed (blue) and dotted (red) lines represent the AFM ordering transition line and the transition line between the ordinary vortex state and the FFLO vortex lattice, respectively. Both of these transitions are of second order in character, while, in the mean field approximation \cite{Adachi}, a first order $H_{c2}$-transition occurs on the black solid curve in the low temperature region shown here. 

As one can see in Fig.\ref{fig:gl} (a) where $\theta=0$, the AFM order appears only in the FFLO region. As indicated in Ref.\cite{Hatakeyama}, the AFM ordering is enhanced by the FFLO order and stabilized by a spatial modulation commensurate with that of the FFLO state. Such appearance of the AFM order is consistent with that found in experiments on CeCoIn$_5$ \cite{recent_NMR,Kenz1,Kenz2}. 

On the other hand, contrary to the experimental fact \cite{recent_NMR,Kenz1,Kenz2}, a {\it nonmagnetic} FFLO region in which the AFM order is unaccompanied inevitably appears at higher temperatures in our calculation shown in Fig.\ref{fig:gl}. However, if AFM fluctuation effects are incorporated beyond the present mean field treatment, this FFLO region is expected to shrink significantly \cite{GL_FFLO2}.

Regarding the relative structure in the direction parallel to the field between the AFM and FFLO orders, the out-of-phase configuration, Fig.2 (b), is found to be realized very rarely in $\theta=0$ case in the present approach using the GL-expansion in $\Delta$. This feature inconsistent with the experimental data \cite{recent_NMR} is found in the next section to be an artifact of the present GL approach and seems to occur because we have kept just one term describing the coupling between the two orders, which is proportional to $|\Delta|^2 \, m^2$, in the present GL approach (see also Ref.\cite{Hatakeyama}). 

Next, the angular dependences of the resulting two orders will be discussed. 
As the field is tilted away from the conducting plane in the uniaxial material, relative contribution of the orbital pair breaking increase, while the paramagnetic effects diminish. Since both of the FFLO and AFM orders have their origin in PPB for the $d$-wave SC pairing state, the field tilt implies that both of the orders are suppressed at higher angles. However, the AFM order has another origin on its suppression due to the field tilt: The Zeeman effect on the AFM ordering occurring from the field component parallel to the AFM moment ${\bf m}$ becomes an origin for suppressing the AFM ordering: As seen in the difference between $j=1$ and $2$ components in eq.(\ref{eq:GL_m2_2}), the nesting property gradually becomes unsatisfactory with increasing the tilt angle. Thus, reduction of the AFM order due to the field-tilt is more remarkable, and a tilt-instability of the AFM order should occur at a lower angle than a threshold angle at which the FFLO phase is lost. Then, a nonmagnetic FFLO phase with no AFM order accompanied needs to exist at lower angles than the FFLO threshold angle. This picture suggested by Fig.\ref{fig:gl} is consistent with the experimental fact \cite{nutron_kaiten,jiwai,jika}.

Further, when the field value $H$ increases, the above-mentioned reduction of the nesting condition due to an increase of the field component parallel to ${\bf m}$ competes with the enhancement of the AFM order due to stronger PPB in larger $H$. Due to this competition, the magnetic field value at which the AFM order is realized at finite temperatures should lie at a lower field than the $H_{c2}(T=0)$. This explains why the dashed (blue) curve in Fig.\ref{fig:gl} (b) shows a field-induced reentry of the AFM order in contrast to that in (a). This result implies that the AFM quantum critical point (QCP), which should occur, in the case of Fig.\ref{fig:gl}, at an angle between 18 and 22 degrees, should also lie at a lower field than $H_{c2}(0)$.

\section{III. Pauli limit}

\hspace{0.3 cm} 
In turn, following the previous work \cite{Hatakeyama}, we will explain our results in the Pauli limit performed in order to examine consequences of the band structure on the HFLT phase of $\rm CeCoIn_{5}$. For this purpose, we start from the conventional tight binding Hamiltonian with a dispersion $\varepsilon({\bf p})$. Broadly, the basic elements in the Hamiltonian are the same as those in the previous section. The only differences are to replace ${\cal H}_{kin}$ in the previous section by 

\begin{eqnarray}
{\cal H}_{\rm 0}'=\sum^{}_{\sigma}\int_{}^{ }d^{3} \mathbf{r}   [\psi^{(\sigma)}]^\dagger(\mathbf{r})
\Bigl[\varepsilon(-{\rm i} \nabla)-\sigma I \Bigr] \psi^{(\sigma)}(\mathbf{r})
\label{h0}
\end{eqnarray}
with
\begin{eqnarray}
\hspace{-14em} 
\psi^{(\sigma)}(\mathbf{r}) &=& \frac{1}{\sqrt{V}}\sum^{}_{\bf{p}}\hat{c}_{\mathbf{p},\sigma}e^{i\mathbf{p}\cdot \mathbf{r}},\nonumber\\
\varepsilon({\bf p}) &=& -2t_{1}(\cos(p_{x}a)+\cos(p_{y}a))-4t_{2}\cos(p_{x}a)\cos(p_{y}a) \nonumber\\
&-& 2t_{3}(\cos(2p_{x}a)+\cos(2p_{y}a))-2t_{4}\cos(p_{z}d)-\mu,
\label{eq:dipersion}
\end{eqnarray}
and to neglect the orbital effect of the magnetic field. 
Hereafter, the Hamiltonian ${\cal H}'={\cal H}_{\rm 0}'+{\cal H}_{\rm SC}+{\cal H}_{\rm AFM}$ is used to obtain the free energy $f_\Delta$, while we will avoid the GL expansion of $f_\Delta$ in the SC order parameter $\Delta$. Due to the neglect of the orbital effect of the magnetic field, the 
SC order parameter in the FFLO phase can be assumed to be homogeneous (uniform), in the real space, in the plane perpendicular to the field so that we assume the form 
$\Delta(\mathbf{r})=|\Delta| \sqrt{2}\cos{(q_{\rm LO} \tilde{y})}$ for the SC order parameter. Further, in our calculation in the Pauli limit, we have assumed the $g$-factor to be isotropic because the anisotropy in the $g$-factor merely leads to a trivial $\theta$-dependent change of the scale of the magnetic field (see also below). 
%$m(\mathbf{r})=|m| \sqrt{2}\cos{(q_{\rm LO} \tilde{y}+ )}$

The normal and anomalous Green's functions in the Matsubara representation are defined as
\begin{eqnarray}
G^{(\sigma)}(\tau; \mathbf{r}_{1},\mathbf{r}_{2}) &=& -\langle T_{\tau} [\psi^{(\sigma)}(\mathbf{r}_{1},\tau) 
[\psi^{(\sigma)}]^\dagger(\mathbf{r}_{2},0)] \rangle ,\nonumber\\
{\overline F}^{(\sigma)}(\tau; \mathbf{r}_{1},\mathbf{r}_{2}) &=& -\langle T_{\tau} [\psi^{(-\sigma)}]^\dagger(\mathbf{r}_{1},\tau) 
[\psi^{(\sigma)}]^\dagger(\mathbf{r}_{2},0)] \rangle ,\nonumber\\
F^{(\sigma)}(\tau; \mathbf{r}_{1},\mathbf{r}_{2}) &=& -\langle T_{\tau} [\psi^{(\sigma)}(\mathbf{r}_{1},\tau) 
\psi^{(-\sigma)}(\mathbf{r}_{2},0)] \rangle ,\nonumber\\
{\overline G}^{(\sigma)}(\tau; \mathbf{r}_{1},\mathbf{r}_{2}) &=& -\langle T_{\tau} [\psi^{(\sigma)}]^\dagger(\mathbf{r}_{1},\tau) 
\psi^{(\sigma)}(\mathbf{r}_{2},0)] \rangle, 
\end{eqnarray}
and the Nambu matrix notation 
\begin{eqnarray}
{\hat G}^{(\sigma)}(\tau; \mathbf{r}_{1},\mathbf{r}_{2}) = \left[ 
\begin{array}{cc}
G^{(\sigma)}(\tau; \mathbf{r}_{1},\mathbf{r}_{2}) & F^{(\sigma)}(\tau; \mathbf{r}_{1},\mathbf{r}_{2}) \\
{\overline F}^{(\sigma)}(\tau; \mathbf{r}_{1},\mathbf{r}_{2}) & {\overline G}^{(-\sigma)}(\tau; \mathbf{r}_{1},\mathbf{r}_{2}) \\
\end{array} 
\right]
\end{eqnarray}
will be used. The Fourier component ${\hat G}^{(\sigma)}_{\varepsilon_{n}}(\mathbf{p};\mathbf{R}) \equiv  
\int d\tau e^{i \varepsilon_{n} \tau} \int d^{3}(\mathbf{r}_{1}-\mathbf{r}_{2})
{\hat G}^{(\sigma)}(\tau; \mathbf{r}_{1},\mathbf{r}_{2})e^{-i\mathbf{p}\cdot (\mathbf{r}_{1}-\mathbf{r}_{2})}$ is represented as 
\begin{eqnarray}
\left[ 
\begin{array}{cc}
i\varepsilon_{n} - \varepsilon({\bf p}+\partial_{\bf R} ) + \sigma I & - \sigma \Delta_{\mathbf{p}}(\mathbf{R}) \\
\sigma \Delta^{\ast}_{\mathbf{p}}(\mathbf{R}) & -i\varepsilon_{n} - \varepsilon({\bf p}+\partial_{\bf R} )  - \sigma I \\
\end{array} 
\right]
{\hat G}^{(\sigma)}_{\varepsilon_{n}}(\mathbf{p};\mathbf{R})
=\hat{1}
\end{eqnarray}
with 
$\mathbf{R}=(\mathbf{r}_{1}+\mathbf{r}_{2})/2$.
The Green's functions are expanded as a power series in the gradient $\nabla_{\mathbf{R}}$ and expressed as
${\hat G}^{(\sigma)}=\hat{G}^{(\sigma)}_{(0)}+{\hat G}^{(\sigma)}_{(2)}+\hat{G}^{(\sigma)}_{(4)}+\cdots$
where
\begin{eqnarray}
{\hat G}^{(\sigma)}_{\varepsilon_{n},\, (0)}(\mathbf{p};\mathbf{R}) &=&
\left[ 
\begin{array}{cc}
i\varepsilon_{n} - \varepsilon(\mathbf{p}) + \sigma I & - \sigma \Delta_{\mathbf{p}}(\mathbf{R}) \\
\sigma \Delta^{\ast}_{\mathbf{p}}(\mathbf{R}) & -i\varepsilon_{n} - \varepsilon(\mathbf{p}) - \sigma I \\
\end{array} 
\right]^{-1} \nonumber\\
&=&
\frac{1}{\varepsilon(\mathbf{p})^{2}-(i\varepsilon_{n}+\sigma I)^{2}+|\Delta_{\mathbf{p}}(\mathbf{R})|^{2}} \left[ 
\begin{array}{cc}
-i\varepsilon_{n} - \varepsilon(\mathbf{p}) - \sigma I & \sigma \Delta_{\mathbf{p}}(\mathbf{R}) \\
- \sigma \Delta^{\ast}_{\mathbf{p}}(\mathbf{R}) & i\varepsilon_{n} - \varepsilon(\mathbf{p}) + \sigma I \\
\end{array} 
\right], \nonumber\\
\label{eq:G0}
\end{eqnarray}
\begin{eqnarray}
{\hat G}^{(\sigma)}_{\varepsilon_n, \, (2)}({\bf p},{\bf R}) &=& {\hat G}^{(\sigma)}_{(0)} \biggl( {\bf v}_{\bf p}\cdot\partial_{\bf R} \biggl(
	{\hat G}^{(\sigma)}_{(0)} {\bf v}_{\bf p}\cdot\partial_{\bf R} {\hat G}^{(\sigma)}_{(0)} \biggr) \biggr) \label{eq:G2}\nonumber \\
{\hat G}^{(\sigma)}_{\varepsilon_n, \, (4)}({\bf p},{\bf R}) &=& {\hat G}^{(\sigma)}_{(0)} \biggl( {\bf v}_{\bf p}\cdot\partial_{\bf R} \biggl(
{\hat G}^{(\sigma)}_{(0)} {\bf v}_{\bf p}\cdot\partial_{\bf R} \biggl( {\hat G}^{(\sigma)}_{(0)} {\bf v}_{\bf p}\cdot\partial_{\bf R} \biggl(
{\hat G}^{(\sigma)}_{(0)} {\bf v}_{\bf p}\cdot\partial_{\bf R} {\hat G}^{(\sigma)}_{(0)} \biggr) \biggr) \biggr) \biggr),\label{eq:G4}  
\end{eqnarray}
\begin{eqnarray}
\partial_{\bf R} = \biggl\{ \begin{array}{cc}
        {\bf \Pi}=-i \nabla_{\bf R} - 2e {\bf A}({\bf R}) & \mbox{for }\Delta({\bf R}) \\ 
        {\bf \Pi}^\dagger = - i \nabla_{\bf R} + 2e {\bf A}({\bf R}) & \mbox{for }\Delta^*({\bf R}) \\ 
        -i\nabla_{\bf R} & \mbox{otherwise}
        \end{array} 
\end{eqnarray}

The mean field free energy associated with the SC order can be constructed in the way \cite{Eilenberger} 
\begin{eqnarray}
f_{\Delta}(q_{\rm LO})
=\biggl\langle 
 \frac{|\Delta(\mathbf{R})|^{2}}{|g|} 
+ \frac{T}{2}
\sum^{\infty}_{\varepsilon_{n}=-\infty} \sum^{}_{\mathbf{p},\sigma} \int_{\varepsilon_{n}}^{\infty s_{\epsilon}}
\!d\omega\  \mathop{\mathrm{Tr}}
\biggl[ i\hat{\sigma}_{z} {\hat G}^{(\sigma)}_{\omega}({\bf p},{\bf R}) \biggr]
\biggr\rangle_{\rm sp} .
\label{eq:fdel}
\end{eqnarray}

By using the relations Eqs.(\ref{eq:G0}) to (\ref{eq:fdel}), the free energy functional is expanded as a power series in the gradient $\nabla_{\mathbf{R}}$ and in the AFM order parameter $m$ and takes the form 
\begin{eqnarray}
f_{\Delta} &=& f_{\Delta}(q_{\rm LO})+f_{m} \nonumber\\
&=& f_{\Delta, (0)}+f_{\Delta, (2)}+f_{\Delta, (4)}+\cdots+f^{(2)}_{m}+f^{(4)}_{m} + \cdots.
\label{eq:F_pertubation}
\end{eqnarray}

The concrete expression of each $f_{\Delta, (n)}$ expressing the $n$-th order term in the gradient is 
\begin{widetext}
\begin{eqnarray}
f_{\Delta,(0)} &=& \biggl\langle \frac{|\Delta(\mathbf{R})|^2}{|g|} - T \sum_{\varepsilon_n>0} \sum_{\bf p}
	\ln \biggl[\frac{(\varepsilon_n^2+[\varepsilon({\bf p})]^2+|\Delta_{\bf p}(\mathbf{R})|^2-I^2)^2+4 \varepsilon_n^2 I^2}{(\varepsilon_n^2+[\varepsilon({\bf p})]^2-I^2)^2+4 \varepsilon_n^2 I^2} \biggr] \biggr\rangle_{\rm sp}, \nonumber \\ 
f_{\Delta, (2)} &=& \biggl\langle T \sum_{\varepsilon_n>0} \sum_{\bf p} \biggl[\frac{a_1^2-b_1^2}{(a_1^2+b_1^2)^2} |{\bf v}_{\bf p}\cdot{\bf \Pi} \Delta_{\bf p}(\mathbf{R})|^2 \nonumber \\
&+& \frac{2}{3} \frac{(2[\varepsilon({\bf p})]^2-\varepsilon_n^2+I^2-|\Delta_{\bf p}(\mathbf{R})|^2) 
(a_1^4-6a_1^2b_1^2+b_1^4) - 4 a_1 b_1^2 (a_1^2-b_1^2) }{(a_1^2+b_1^2)^4} ({\bf v}_{\bf p}\cdot\nabla |\Delta_{\bf p}(\mathbf{R})|^2)^2 \biggr] \biggr\rangle_{\rm sp}, 
\nonumber \\
f_{\Delta, (4)} &\simeq& \biggl\langle \frac{2T}{3} \sum_{\varepsilon_n>0} \sum_{\bf p} 
\biggl[ \frac{(2 [\varepsilon({\bf p})]^2-\varepsilon_n^2+I^2-|\Delta_{\bf p}(\mathbf{R})|^2) 
(a_1^4-6a_1^2b_1^2+b_1^4) - 4 a_1 b_1^2 (a_1^2 - b_1^2)}{(a_1^2+b_1^2)^4}|({\bf v}_{\bf p}\cdot{\bf \Pi})^2 \Delta_{\bf p}(\mathbf{R})|^2 \biggr] \biggr\rangle_{\rm sp},
\label{FSCgrad}
\end{eqnarray}
\end{widetext}
respectively, where 
$a_1 = [\varepsilon({\bf p})]^2+\varepsilon_n^2+|\Delta_{\bf p}(\mathbf{R})|^2-I^2$, and $b_1 = 2 \varepsilon_n I$. 
On the other hand, the terms $f^{(2)}_{m}$ and $f^{(4)}_{m}$ describing the mean field AFM ordering are expressed as
\begin{eqnarray}
f^{(2)}_{m} &=& \biggl\langle
\biggl[
\frac{1}{U} 
+ \frac{T}{2}
\sum^{\infty}_{\varepsilon_{n}=-\infty} \sum^{}_{\mathbf{p},\sigma} 
{\rm Tr}
\biggl(
\sum^{2}_{j=1}
A_{j}
\hat{a}_{j} {\hat G}^{(\sigma)}_{\varepsilon_n, \, (0)}({\bf p};{\bf R})
\hat{b}_{j} {\hat G}^{(\alpha_{j})}_{\varepsilon_n, \, (0)}({\bf p}+\mathbf{Q}_{0}+\mathbf{q};{\bf R})
\biggr)
\biggr]
|m(\mathbf{R})|^{2} \biggr\rangle_{\rm sp}, \nonumber \\ 
f^{(4)}_{m} &=& \biggl\langle
\frac{T}{4}
\sum^{\infty}_{\varepsilon_{n}=-\infty} \sum^{}_{\mathbf{p},\sigma} 
{\rm Tr}
\biggl(
\sum^{5}_{j=1}
A''_{j} 
\hat{a}''_{j} {\hat G}^{(\sigma)}_{\varepsilon_n, \, (0)}({\bf p};{\bf R})
\hat{b}''_{j} {\hat G}^{(\alpha''_{j})}_{\varepsilon_n, \, (0)}({\bf p}+\mathbf{Q}_{0}+\mathbf{q};{\bf R}) \nonumber \\ 
&\times& \hat{c}''_{j} {\hat G}^{(\beta''_{j})}_{\varepsilon_n, \, (0)}({\bf p};{\bf R})
\hat{d}''_{j} {\hat G}^{(\gamma''_{j})}_{\varepsilon_n, \, (0)}({\bf p}+\mathbf{Q}_{0}+\mathbf{q};{\bf R})
\biggr)
|m(\mathbf{R})|^{4}
\biggr\rangle_{\rm sp},
\label{eq:pauli_AFM}
\end{eqnarray}

and the coefficients $A_{j}$, $\alpha_{j}$, $\hat{a}_{j}$, $\hat{b}_{j}$, $A''_{j}$, $\alpha''_{j}$, $ \beta''_{j}$, $\gamma''_{j}$
, $\hat{a}''_{j}$, $\hat{b}''_{j}$, $\hat{c}''_{j}$, and $\hat{d}''_{j}$ are represented in Table \ref{tb:table4} and Table \ref{tb:table5}.

\begin{table}[htb]
%\begin{center}
\caption{Coefficients $\alpha_{j}$, $A_{j}$, $B_{j}$ in Eq. (\ref{eq:pauli_AFM}) }
  \begin{tabular}{ccccc} \hline
    $j$ & $\alpha_{j}$ & $A_{j}$ & $\hat{a}_{j}$ & $\hat{b}_{j}$ \\ \hline
    $1$ & $-\sigma$ & $\cos^{2}\theta$ & $\hat{1}$ & $\hat{1}$ \\
    $2$ & $\sigma$ & $\sin^{2}\theta$ & $\hat{\sigma}_{z}$ & $\hat{\sigma}_{z}$ \\ \hline
  \end{tabular}
  \label{tb:table4}
%  \end{center}
\end{table}

\begin{table}[htb]
%\begin{center}
\caption{Coefficients $\alpha''_{j}$, $ \beta''_{j}$, $\gamma''_{j}$, $A''_{j}$, $\hat{a}''_{j}$, $\hat{b}''_{j}$, $\hat{c}''_{j}$, $\hat{d}''_{j}$
 in Eq. (\ref{eq:pauli_AFM}) }
  \begin{tabular}{ccccccccc} \hline
    $j$ & $\alpha''_{j}$ & $\beta''_{j}$ & $\gamma''_{j}$ & $A''_{j}$ & $\hat{a}''_{j}$ & $\hat{b}''_{j}$ & $\hat{c}''_{j}$ & $\hat{d}''_{j}$ \\ \hline
    $1$ & $-\sigma$ & $\sigma$ & $-\sigma$ & $\cos^{4}\theta$ & $\hat{1}$ & $\hat{1}$ & $\hat{1}$ & $\hat{1}$ \\
    $2$ & $\sigma$ & $\sigma$ & $\sigma$ & $\sin^{4}\theta$ & $\hat{\sigma}_{z}$ & $\hat{\sigma}_{z}$ & $\hat{\sigma}_{z}$ & $\hat{\sigma}_{z}$ \\
    $3$ & $\sigma$ & $\sigma$ & $-\sigma$ & $2\cos^{2}\theta \sin^{2}\theta$ & $\hat{1}$ & $\hat{\sigma}_{z}$ & $\hat{\sigma}_{z}$ & $\hat{1}$ \\
    $4$ & $\sigma$ & $-\sigma$ & $\sigma$ & $2\cos^{2}\theta \sin^{2}\theta$ & $\hat{\sigma}_{z}$ & $\hat{\sigma}_{z}$ & $\hat{1}$ & $\hat{1}$ \\
    $5$ & $\sigma$ & $-\sigma$ & $-\sigma$ & $-2\cos^{2}\theta \sin^{2}\theta$ & $\hat{1}$ & $\hat{\sigma}_{z}$ & $\hat{1}$ & $\hat{\sigma}_{z}$ \\ \hline
\end{tabular}
\label{tb:table5}
%\end{center}
\end{table}

Finally, the concrete expressions of the free energy $f^{(2)}_{m}$ and $f^{(4)}_{m}$ in Eq. (\ref{eq:pauli_AFM}) are represented by 

\begin{eqnarray}
f^{(2)}_{m} &=& \biggl\langle
\biggl[
\frac{1}{U} 
+ 2T
\sum^{}_{\varepsilon_{n}>0} \sum^{}_{\mathbf{p}} 
\biggl(
\sum^{2}_{j=1}
A_{j} V^{(2)}_{m, \, j}(\mathbf{p},\mathbf{q};{\bf R})
\biggr)
\biggr]
|m(\mathbf{R})|^{2}
\biggr\rangle_{\rm sp}, \nonumber \\ 
f^{(4)}_{m} &=& \biggl\langle
T
\sum^{}_{\varepsilon_{n}>0} \sum^{}_{\mathbf{p}}
\biggl(
\sum^{5}_{j=1}
A''_{j} V^{(4)}_{m, \, j}(\mathbf{p},\mathbf{q};{\bf R})
\biggr)
|m(\mathbf{R})|^{4}
\biggr\rangle_{\rm sp},
\end{eqnarray}
where the details of $V_{m, \, j}^{(2)}$ and $V_{m, \, j}^{(4)}$ will be given in Appendix. 

%%%%%%%%%%%%%%%%%%%%%%%%%%%%%%%%%%%%%%%%%%%%%%%%%%%%
\begin{figure}[b]
\scalebox{0.6}[0.6]{\includegraphics{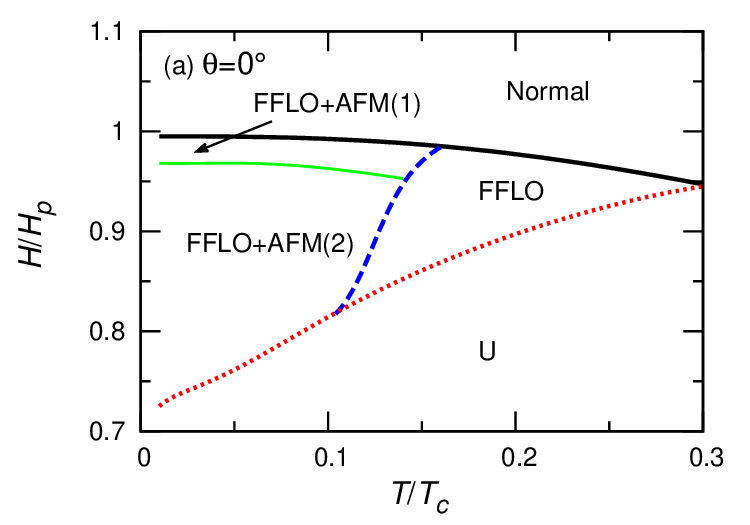}}
\scalebox{0.6}[0.6]{\includegraphics{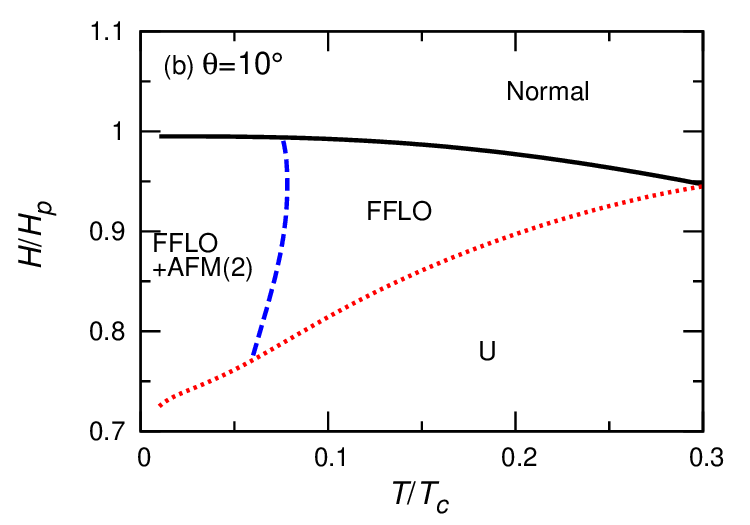}}
\scalebox{0.6}[0.6]{\includegraphics{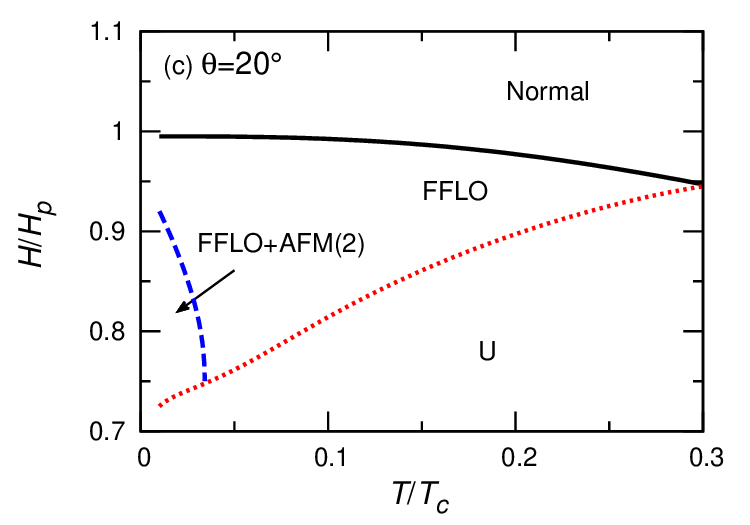}}
\scalebox{0.6}[0.6]{\includegraphics{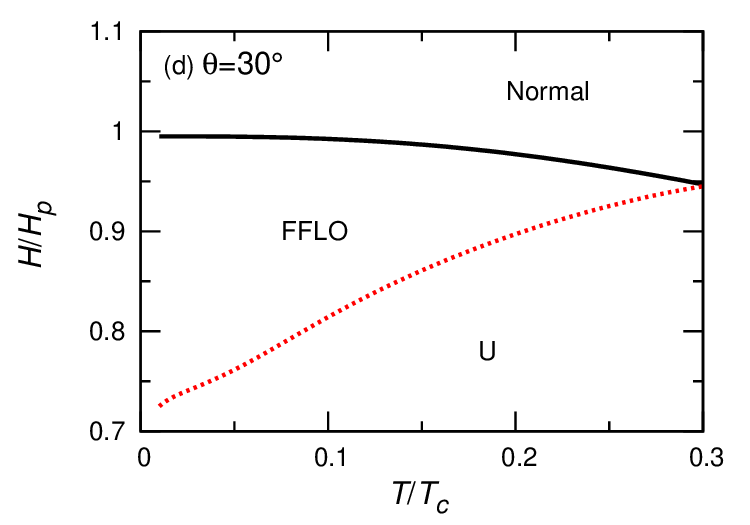}}
\caption{Angular dependence of the $H\mathchar`-T$ phase diagrams in the Pauli limit. In this calculation, the parameter values $T_{c}/U=0.01597$, $t_{1}/T_{c}=15, t_{2}/t_{1}=-1.5, t_{3}/t_{1}=0.65, t_{4}/t_{1}=0.5$, and $\mu/t_{1}=1.85$ are used. In each of the four figures, the thin solid (green) line in (a) denotes a structural transition on the relative configuration between the spatially modulated AFM order and the FFLO one (see the text), while the remaining curves are defined in the same manner as those in Fig.3. The uniform SC (U) phase corresponds to the Abrikosov lattice (A) in Fig.2, i.e., the case including the orbital pair-breaking.}
  \label{fig:p_phase_diagram}
\end{figure}
%%%%%%%%%%%%%%%%%%%%%%%%%%%%%%%%%%%%%%%%%%%%%%%%%%%

\subsection{\it Results}

Now, numerically obtained results on the phase diagram will be explained below. Figure \ref{fig:p_phase_diagram} (a)`(d) shows field v.s. temperature ($H$-$T$) phase diagrams including the PPB-induced AFM and FFLO orders obtained at some fixed angles and in the Pauli limit with no orbital pair-breaking. In each figure, the thick solid (black) curve denote the first order $H_{c2}$ transition line, and the dotted (red) and dashed (blue) one denotes the continuous transition lines to the FFLO and AFM ordered phases, respectively. The green dashed one in (a) denotes a transition line separating the two different AFM-FFLO coupled structures shown in Fig.2 from each other. 
As examined in Ref.\cite{Hatakeyama}, in the parallel fields case shown in Fig.\ref{fig:p_phase_diagram} (a), AFM order tends to appear just in the FFLO phase because AFM order is induced by PPB effects and, in addition, is stabilized by taking a spatial modulation commensurate to the FFLO modulation parallel to the magnetic field. However, the relative phase between the spatial modulations of AFM and FFLO may be changed depending upon the field strength. As Fig.4 (a) shows, the AFM order favors coexistence (i.e, the structure of Fig.2 (a)) with the nonvanishing SC order in real space in higher fields, which is a feature consistent with the NMR data \cite{recent_NMR}, and a structural transition on the green line is expected to occur with decreasing the field in the HFLT phase. 

The resulting angular dependence of the AFM ordered region in this Pauli-limited model is similar to that in the preceding section. With tilting the field away from the superconducting plane, the nesting condition for the antiferromagnetism becomes gradually unsatisfactory, because the magnetic field component parallel to the AFM moment ${\bf m} \parallel c$ \cite{Kenz1,Kenz2,nutron_kaiten} is increased by the tilt so that the incommensurate AFM wavevectors for each spin components do not coincide with each other when $\theta \neq 0$. Therefore, as shown in the Fig.\ref{fig:p_phase_diagram} (b), AFM order tends to be suppressed by the tilt even without the orbital pair-breaking. However, the primary origin of this AFM order in the SC order, i.e., the PPB effect, is enhanced with increasing the field. Due to a competition between this field-induced enhancement of PPB and the above-mentioned less complete nesting condition due to the field-tilt, the resulting AFM phase is pushed down to lowr fields within the FFLO phase. Thus, the pure (nonmagnetic) FFLO state appears between the AFM ordered region and the $H_{c2}$-line. 
In contrast to the GL approach in the preceding section, we have not assumed an anisotropy of the $g$-factor. For this reason, the resulting $H_{c2}$-curve in Fig.\ref{fig:p_phase_diagram} is independent of the angle. A more realistic angular dependence of the $H$-$T$ phase diagram in the Pauli limit is trivially obtained by properly changing the scale of the ordinate of the figures (b) to (d). It is remarkable that, nevertheless, the real space structure of the AFM order is highly sensitive to the field-tilt: The in-phase structure, FFLO $+$ AFM(1) in the figure (a), (i.e., Fig.2(a)), seen in the parallel field case is lost by a small tilt of the field, and the resulting AFM order for $\theta \neq 0$ basically takes the out-of-phase structure (FFLO $+$ AFM(2)) defined in Fig.2 (b). It seems that a weaker AFM ordering leads to the out-of-phase structure in which the AFM order appears in the region where the SC energy gap vanishes. 

We need to stress that the out-of-phase configuration of the AFM order is primarily realized in the present case in contrast to that in the preceding section based on the GL expansion. As the only coupling between the two orders in the GL expansion, just the O($|\Delta|^2 \, m^2$) term has been incorporated in the preceding section, while, up to O($m^2$), the AFM order couples to the SC one in all orders in $\Delta$ in this Pauli limit. Due only to this difference in the theoretical treatment, the real space structure of the AFM order has been changed between the two theoretical approaches. It seems that the GL expansion will not be sufficient to study the detailed structures of an ordered phase. 

In the present Pauli-limited model, FFLO order is not suppressed sufficiently by the tilt. This seemingly inconsistent behavior with experimental facts is a result of our neglect in this model of the orbital pair-breaking effect. As seen in the preceding section, inclusion of the orbital pair-breaking effect recovers a tilt-induced reduction \cite{jiwai,jika,Kumagaionly} of the FFLO ordered region. 

\section{IV. Summary}

In the present paper, we have investigated, within the mean field approximation, how the high-field low-temperature (HFLT) SC phase found in CeCoIn$_5$ in the parallel field configuration is changed by rotating (or, tilting) the field direction from the basal plane. Since it is difficult at present to perform a complete analysis incorporating various features of the ordered states due to PPB effects on the same footing, we have examined two models separately to obtain basic knowledges on the present issue. In one model, we have incorporated the orbital pairing breaking, i.e., the presence of the vortices, while, instead, the GL expansion in the SC order parameter $\Delta$ has been assumed. Due to the use of the GL expansion in $\Delta$, the relative structure in real space between the resulting AFM order and the FFLO modulation is found not to become consistent with the recent experimental data \cite{Kumagaionly}. Instead, we have found such a feature consistent with the experimental data \cite{jiwai,jika} that the region of the HFLT phase in the $H$-$T$ phase diagram, identified with the FFLO state itself, is significantly reduced with tilting the field, because the orbital pair-breaking included in this GL approach becomes more important, due to the uniaxial crystalline anisotropy, with tilting the field. 

On the other hand, to improve our understanding on the details of the resulting AFM order in real space, we have also examined the Pauli-limited model in which $\Delta$ is uniform in the plane perpendicular to the field due to the neglect of the orbital pair-breaking, while the GL expansion in $\Delta$ is not assumed. In this model, the FFLO ordering is overestimated, and the correct angular dependence of the FFLO region is not obtained. Instead, reflecting higher order couplings between the AFM and SC order parameters which are absent in the GL approach in sec.II, both of the two different configurations, illustrated in Fig.2, of the AFM and FFLO structures parallel to the field are obtained in the parallel field configuration with a field dependence qualitatively consistent with the data \cite{recent_NMR}. 

Main results in the present work is the angular dependence of the AFM ordered region in the FFLO phase found {\it commonly} both in our two models in the preceding two sections. Originally, an increase of the magnetic field enhances PPB effects, i.e., the present AFM ordering, while the $c$-axis component of the magnetic field induced by tilting the field-direction from the basal plane makes the nesting condition for the AFM unsatisfactory as far as the AFM moment is parallel to the $c$-axis \cite{Kenz1,Kenz2,nutron_kaiten}. These two competitive roles of the tilted magnetic field for the AFM ordering shift the field range, in which the AFM ordering is maximal, to lower fields than $H_{c2}(0)$. Consequently, when tilting the field, the AFM order is first lost in the {\it high} field range just below $H_{c2}(T)$-line. This strikingly coincides with the feature seen through a NMR measurement \cite{Kumagaionly} in the angular dependence of the AFM ordered region in the HFLT phase of CeCoIn$_5$. In addition, this type of reduction of the AFM order suggests that, at higher tilt angles, a remaining AFM quantum critical point (AFM-QCP) should lie at a slightly lower field than $H_{c2}(0)$. This expected position of the AFM-QCP seems to be consistent with the experimental facts \cite{critical_Ce2,Kasahara} in ${\bf H} \parallel c$ which show an AFM-QCP not coinciding with $H_{c2}(0)$ but lying clearly below it. 
Further, we have found, in the Pauli limit where the GL expansion in $\Delta$ is not used, that inclusion of an effect suppressing the AFM ordering in the FFLO, such as the field-tilt, state results in formation of the AFM order close to the FFLO nodal planes. It means that, as seen in Fig.\ref{fig:p_phase_diagram} (b)-(d), a field tilt from the in-plane field configuration results in the structure illustrated in Fig.2 (b) in contrast to the high field behavior in the in-plane field configuration seen in Fig.\ref{fig:p_phase_diagram} (a). This change on the AFM ordering by the field-tilt has also been found in recent NMR data on the angular dependence of the HFLT phase of CeCoIn$_5$ \cite{Kumagaionly}. 

The present theory on the high field AFM ordering in superconductors with strong PPB is the first study on the angular ($\theta$-) dependence of the HFLT SC phase found in CeCoIn$_5$. Further, even if focusing on the parallel field case with $\theta=0$, this is different from other works \cite{Yanase1,Machida,Ilya}. First of all, our works take account of correlation between the AFM and the FFLO orders, while just the AFM order has been considered as a consequence of PPB elsewhere \cite{Machida,Ilya}. We stress that the region with no AFM order in the HFLT SC phase has been realized at least by the field tilt \cite{Kumagaionly}. Second of all, our works explain coexistence of the AFM order and the nonvanishing SC order parameter in real space which has been seen in the higher half of the field range of the HFLT phase \cite{recent_NMR}, while the picture in Ref.\cite{Yanase1} requires the resulting AFM order, as in Fig.2 (b), in any case  to localize close to the FFLO nodal plane. 

Finally, we should comment on the local approximation used to simplify our treatment. Effects of the FFLO order on the AFM order are incorporated even in this approximation. However, as explained in the text of sec.II, in this local approximation, the couplings between the AFM and the SC orders of higher orders in the FFLO wavenumber $|q_{LO}|$ are neglected. This higher order coupling would describe possible effects of the AFM order on the FFLO order and thus, on the $H_{c2}(T)$-curve at low temperatures. Further study including these higher order couplings will be reported elsewhere. 

We thank K. Aoyama for his contribution on early stage of the present work and discussions and Y. Hatakeyama for discussions. This work is supported by Grant-in-Aid for Scientific Research [No.25400368] from MEXT, Japan. 
 
\section{Appendix}

\hspace{0.3 cm} The GL free energy functional in Eq.(\ref{eq:sc_part}) can be obtained simply by extending the previous analysis \cite{Adachi,GL_FFLO1} to the present tilted case, and its each term is given by 
\begin{eqnarray}
f^{(2,0)}_\Delta %\frac{{\cal F'}_{\Delta^{2},q_{\rm LO}^{0}}}{N(0)}
&=& N(0) T_{c}^{2} %\frac{N(0)}{2!}
\left[\frac{1}{2} \ln{\left( \frac{H}{H_{\rm orb}^{\rm 2D}(0)}\right)} + 
\int_{0}^{\infty} d\rho \biggl(\frac{1}{\rho} \exp{\left( -\frac{\pi^2 \xi_0^2}{r_H^2} \rho^{2} \right) }
- f(\rho,\rho) 
\biggl\langle
|w_{\mathbf{p}}|^{2}
 \exp(-\frac{1}{2}|\mu|^{2}\rho^{2} )
\biggr\rangle_{\rm FS} \biggr)
 \right]
\biggl( \frac{|\Delta|}{T_{c}} \biggr)^{2}  , \nonumber \\ 
f^{(2,2)}_\Delta %\frac{{\cal F'}_{\Delta^{2},q_{\rm LO}^{2}}}{N(0)} 
&=& \frac{N(0)T_{c}^{2}}{2} \int_{0}^{\infty} d\rho  f(\rho,\rho) \rho^{2}
\biggl\langle
|w_{\mathbf{p}}|^{2} 
\biggl( \frac{ \tilde{v}_{\tilde{p},\tilde{y}} }{v_{\rm F}} \biggr)^{2}
 \exp(-\frac{1}{2}|\mu|^{2}\rho^{2} )
\biggr\rangle_{\rm FS} 
\biggl( \frac{|\Delta|}{T_{c}} \biggr)^{2} 
\biggl( \frac{v_{\rm F}}{T_{c}} \biggr)^{2} , \nonumber \\ 
%\biggl( \frac{|\Delta|}{T_{c}} \biggr)^{2} \biggl( \frac{v_{\rm F}q_{\rm LO}}{T_{c}} \biggr)^{2}
f^{(2,4)}_\Delta %\frac{{\cal F'}_{\Delta^{2}, q_{\rm LO}^{4}}}{N(0)} 
&=& -\frac{N(0)T_{c}^{2}}{24} \int_{0}^{\infty} d\rho  f(\rho,\rho) \rho^{4}
\biggl\langle
|w_{\mathbf{p}}|^{2} 
\biggl( \frac{ \tilde{v}_{\tilde{p},\tilde{y}} }{v_{\rm F}} \biggr)^{4}
 \exp(-\frac{1}{2}|\mu|^{2}\rho^{2} )
\biggr\rangle_{\rm FS} 
\biggl( \frac{|\Delta|}{T_{c}} \biggr)^{2} 
\biggl( \frac{v_{\rm F}}{T_{c}} \biggr)^{4} , \nonumber \\ 
f^{(4,0)}_\Delta %\frac{{\cal F'}_{\Delta^{4}, q_{\rm LO}^{0}}}{N(0)}
&=& \frac{3N(0)T_{c}^{2}}{2}\int_{0}^{\infty} \prod^{3}_{i=1}d\rho_{i} f\biggl(\sum^{3}_{i=1} \rho_{i}, \sum^{3}_{i=1} \rho_{i} \biggr)
\biggl\langle
|w_{\mathbf{p}}|^{4}  \exp{\left[-\frac{1}{2}  \left( R_{14} -\frac{1}{2}R_{24} \right) \right]  }
\cos(I_{4}) \biggr\rangle_{\rm FS} 
\biggl( \frac{|\Delta|}{T_{c}} \biggr)^{4} , \nonumber \\ 
f^{(4,2)}_\Delta %\frac{{\cal F'}_{\Delta^{4}, q_{\rm LO}^{2}}}{N(0)}
&=& -\frac{3N(0)T_{c}^{2}}{4}\int_{0}^{\infty} \prod^{3}_{i=1}d\rho_{i} f\biggl(\sum^{3}_{i=1} \rho_{i}, \sum^{3}_{i=1} \rho_{i} \biggr)
 \biggl[ \sum^{3}_{i=1}\rho_{i}^{2} - \frac{1}{3} \sum^{}_{i\neq j}(-1)^{i+j}\rho_{i}\rho_{j} \biggr] , \nonumber \\ 
&\times& \biggl\langle
|w_{\mathbf{p}}|^{4} \biggl( \frac{ \tilde{v}_{\tilde{p},\tilde{y}} }{v_{\rm F}} \biggr)^{2}
 \exp{\left[-\frac{1}{2}  \left( R_{14} -\frac{1}{2}R_{24} \right) \right]  }
\cos(I_{4}) \biggr\rangle_{\rm FS} 
\biggl( \frac{|\Delta|}{T_{c}} \biggr)^{4} 
\biggl( \frac{v_{\rm F}}{T_{c}} \biggr)^{2} , \nonumber \\ 
f^{(4,4)}_\Delta %\frac{{\cal F'}_{\Delta^{4}, q_{\rm LO}^{4}}}{N(0)}
&=& \frac{N(0)T_{c}^{2}}{16}\int_{0}^{\infty} \prod^{3}_{i=1}d\rho_{i} f\biggl(\sum^{3}_{i=1} \rho_{i}, \sum^{3}_{i=1} \rho_{i} \biggr) \Bigl[ \sum^{3}_{i=1}\rho_{i}^{4} + \sum^{}_{i\neq j}\Bigl[3\rho_{i}^{2}\rho_{j}^{2} - 2(-1)^{i+j}\rho_{i}\rho_{j}(\rho_{6-i-j})^{2} -\frac{3}{4}(-1)^{i+j} \rho_{i}\rho_{j}^{3}  \Bigr] \Bigr] , \nonumber \\ 
&\times& \biggl\langle
|w_{\mathbf{p}}|^{4} \biggl( \frac{ \tilde{v}_{\tilde{p},\tilde{y}} }{v_{\rm F}} \biggr)^{4}
 \exp{\left[-\frac{1}{2}  \left( R_{14} -\frac{1}{2}R_{24} \right) \right]  }
\cos(I_{4}) \biggr\rangle_{\rm FS} 
\biggl( \frac{|\Delta|}{T_{c}} \biggr)^{4} 
\biggl( \frac{v_{\rm F}}{T_{c}} \biggr)^{4}  , \nonumber \\ 
f^{(6)}_\Delta %\frac{{\cal F'}_{\Delta^{6}, q_{\rm LO}^{0}}}{N(0)}
&=& -\frac{5N(0)T_{c}^{2}}{2}\int_{0}^{\infty} \prod^{5}_{i=1}d\rho_{i} f\biggl(\sum^{5}_{i=1} \rho_{i}, \sum^{5}_{i=1} \rho_{i} \biggr)
 \biggl\langle
|w_{\mathbf{p}}|^{6}  \exp{\left[-\frac{1}{2}  \left( R_{16}+R_{26} \right) \right]  }
\cos(I_{6})
\biggr\rangle_{\rm FS} 
\biggl( \frac{|\Delta|}{T_{c}} \biggr)^{6}  ,
\end{eqnarray}
where $\xi_0$ is the in-plane coherence length, 
\begin{eqnarray}
R_{14} & = &|\mu|^{2}\left[\sum^{3}_{i=1}\rho_{i}^{2}+\rho_{2}(\rho_{1}+\rho_{3})\right]\nonumber\\
R_{24} & = &{\rm Re}(\mu^{2})\left[\rho_{2}^{2}+(\rho_{1}-\rho_{3})^{2}\right]\nonumber\\
I_{4} & = &\frac{1}{4} {\rm Im} (\mu^{2})\left[-\rho_{2}^{2}+(\rho_{1}-\rho_{3})^{2}\right] \nonumber\\
R_{16} & = &|\mu|^{2}\left(e_{1}+e_{2}+e_{3}+\frac{2}{3}e_{4}e_{5}\right) \nonumber\\
R_{26} & = &{\rm Re}(\mu^{2})\left[e_{1}+e_{2}+e_{3}-\frac{e_{4}^{2}+e_{5}^{2}}{3}
-\frac{2}{3} ( e_{6}+e_{7}+e_{8}+e_{9}) 
\right] \nonumber\\
I_{6} & = &\frac{1}{4} {\rm Im} (\mu^{2}) \left[e_{1}+e_{2}-e_{3}-\frac{e_{4}^{2}-e_{5}^{2}}{3}
-\frac{2}{3} ( e_{6}+e_{7}-e_{8}-e_{9}) 
\right] \nonumber\\
e_{1} & = &(\rho_{3}+\rho_{5})^{2}+(\rho_{3}+\rho_{4})^{2}\nonumber\\
e_{2} & = &(\rho_{1}+\rho_{4}+\rho_{5})^{2}\nonumber\\
e_{3} & = &\rho_{3}^{2}+\rho_{4}^{2}+(\rho_{2}-\rho_{5})^{2}\nonumber\\
e_{4} & = &\rho_{1}+2(\rho_{3}+\rho_{4}+\rho_{5})\nonumber\\
e_{5} & = &\rho_{2}-\rho_{3}-\rho_{4}-\rho_{5}\nonumber\\
e_{6} & = &(\rho_{4}-\rho_{5})^{2}+(\rho_{1}-\rho_{3}+\rho_{5})^{2}\nonumber\\
e_{7} & = &(\rho_{1}-\rho_{3}+\rho_{4})^{2}\nonumber\\
e_{8} & = &(\rho_{3}-\rho_{4})^{2}+(\rho_{2}+\rho_{3}-\rho_{5})^{2}\nonumber\\
e_{9} & = &(\rho_{2}+\rho_{4}-\rho_{5})^{2}.
\nonumber\\
\end{eqnarray}

\begin{eqnarray*}
\hspace{0em}
V^{(2)}_{m, \, 1}(\mathbf{p},\mathbf{q};{\bf R})
&=& - \frac{2}{ a_{1}^{2}+b_{1}^{2} }
\bigl[ 
c_{1} [
\varepsilon_{n}^{2}+I^{2} - \varepsilon(\mathbf{p})\varepsilon(\mathbf{p}+\mathbf{Q}_{0} + \mathbf{q}) 
- \Delta_{\mathbf{p}}(\mathbf{R}) \Delta^{\ast}_{\mathbf{p}+\mathbf{Q}_{0}+\mathbf{q}}(\mathbf{R})] 
\bigr]
, \nonumber \\ 
V^{(2)}_{m, \, 2}(\mathbf{p},\mathbf{q};{\bf R})
&=& - \frac{2}{ a_{2}^{2}+b_{1}^{2} }
\bigl[ 
c_{2}
[
\varepsilon_{n}^{2}-I^{2}-\varepsilon(\mathbf{p})\varepsilon(\mathbf{p}+\mathbf{Q}_{0}+\mathbf{q}) 
- \Delta_{\mathbf{p}}(\mathbf{R}) \Delta^{\ast}_{\mathbf{p}+\mathbf{Q}_{0}+\mathbf{q}}(\mathbf{R})]
+ d_{2} b_{1} 
\bigr]
 , \nonumber \\ 
V^{(4)}_{m, \, 1}(\mathbf{p},\mathbf{q};{\bf R})
&=& \frac{2}{ ( c_{1}^{2}+d_{1}^{2} )^{2} }
\bigl[ ( c_{1}^{2}-d_{1}^{2} ) [ ( \varepsilon_{n}^{2}+I^{2}-\varepsilon(\mathbf{p})\varepsilon(\mathbf{p}+\mathbf{Q}_{0} + \mathbf{q}) 
- \Delta_{\mathbf{p}}(\mathbf{R}) \Delta^{\ast}_{\mathbf{p}+\mathbf{Q}_{0}+\mathbf{q}}(\mathbf{R}) )^{2}  \nonumber \\ 
&-& \varepsilon_{n}^{2} (  ( \varepsilon(\mathbf{p}) + \varepsilon(\mathbf{p}+\mathbf{Q}_{0} + \mathbf{q}) )^{2} 
+ |\Delta_{\mathbf{p}}(\mathbf{R}) - \Delta_{\mathbf{p}+\mathbf{Q}_{0}+\mathbf{q}}(\mathbf{R})  |^{2}    )
+ I^{2} (  ( \varepsilon(\mathbf{p}) - \varepsilon(\mathbf{p}+\mathbf{Q}_{0} + \mathbf{q}) )^{2}  \nonumber \\ 
&+& |\Delta_{\mathbf{p}}(\mathbf{R}) - \Delta_{\mathbf{p}+\mathbf{Q}_{0}+\mathbf{q}}  |^{2}) 
- |\Delta_{\mathbf{p}}(\mathbf{R}) \varepsilon(\mathbf{p}+\mathbf{Q}_{0} + \mathbf{q})
- \Delta_{\mathbf{p}+\mathbf{Q}_{0}+\mathbf{q}}(\mathbf{R}) \varepsilon(\mathbf{p}) |^{2}
]
+ 2 c_{1} d_{1}^{2} b_{1}
\bigr]
, \nonumber \\ 
V^{(4)}_{m, \, 2}(\mathbf{p},\mathbf{q};{\bf R})
&=& \frac{2}{ ( c_{2}^{2}+d_{2}^{2} )^{2} }
\bigl[ ( c_{2}^{2}-d_{2}^{2} ) [ ( \varepsilon_{n}^{2}-I^{2}-\varepsilon(\mathbf{p})\varepsilon(\mathbf{p}+\mathbf{Q}_{0} + \mathbf{q}) -
\Delta_{\mathbf{p}}(\mathbf{R}) \Delta^{\ast}_{\mathbf{p}+\mathbf{Q}_{0}+\mathbf{q}}(\mathbf{R}) )^{2} - b_{1}^{2}  \nonumber \\ 
&-& (\varepsilon_{n}^{2} - I^{2} )( ( \varepsilon(\mathbf{p}) + \varepsilon(\mathbf{p}+\mathbf{Q}_{0} + \mathbf{q}) )^{2} 
+ |\Delta_{\mathbf{p}}(\mathbf{R}) + \Delta_{\mathbf{p}+\mathbf{Q}_{0}+\mathbf{q}}(\mathbf{R})  |^{2}    )
- |\Delta_{\mathbf{p}}(\mathbf{R}) \varepsilon(\mathbf{p}+\mathbf{Q}_{0} + \mathbf{q}) \nonumber \\ 
&-& \Delta_{\mathbf{p}+\mathbf{Q}_{0}+\mathbf{q}}(\mathbf{R}) \varepsilon(\mathbf{p}) |^{2} ]  
- 2 c_{1} d_{1} b_{1} 
[
( \varepsilon(\mathbf{p}) + \varepsilon(\mathbf{p}+\mathbf{Q}_{0} + \mathbf{q}) )^{2} 
- 2(\varepsilon_{n}^{2}-I^{2}- \varepsilon(\mathbf{p}) \varepsilon(\mathbf{p}+\mathbf{Q}_{0} + \mathbf{q})  ) \nonumber \\ 
&+& 2 \Delta_{\mathbf{p}}(\mathbf{R}) \Delta^{\ast}_{\mathbf{p}+\mathbf{Q}_{0}+\mathbf{q}}(\mathbf{R})  
+  |\Delta_{\mathbf{p}}(\mathbf{R}) + \Delta_{\mathbf{p}+\mathbf{Q}_{0}+\mathbf{q}}(\mathbf{R})  |^{2} 
]
\bigr]
, \nonumber \\ 
V^{(4)}_{m, \, 3}(\mathbf{p},\mathbf{q};{\bf R})
&=& \frac{2}{( a_{2}^{2}+b_{1}^{2} )( c_{3}^{2}+d_{3}^{2} )}
\bigl[ 
c_{3}[ 
( \varepsilon_{n}^{2}+I^{2}-\varepsilon(\mathbf{p})\varepsilon(\mathbf{p}+\mathbf{Q}_{0} + \mathbf{q}) 
- \Delta_{\mathbf{p}}(\mathbf{R}) \Delta^{\ast}_{\mathbf{p}+\mathbf{Q}_{0}+\mathbf{q}}(\mathbf{R})  )  \nonumber \\ 
&\times& (  \varepsilon_{n}^{2}-I^{2}-\varepsilon(\mathbf{p})\varepsilon(\mathbf{p}+\mathbf{Q}_{0} + \mathbf{q}) 
- \Delta_{\mathbf{p}}(\mathbf{R}) \Delta^{\ast}_{\mathbf{p}+\mathbf{Q}_{0}+\mathbf{q}}(\mathbf{R}) ) 
- \varepsilon_{n}^{2} (     (\varepsilon(\mathbf{p}) + \varepsilon(\mathbf{p}+\mathbf{Q}_{0} + \mathbf{q}) )^{2} \nonumber \\ 
&+& | \Delta_{\mathbf{p}}(\mathbf{R}) + \Delta_{\mathbf{p}+\mathbf{Q}_{0}+\mathbf{q}}(\mathbf{R})  |^{2}    )
- I^{2} (\varepsilon(\mathbf{p})^{2} - \varepsilon(\mathbf{p}+\mathbf{Q}_{0} + \mathbf{q})^{2}  
+ | \Delta_{\mathbf{p}}(\mathbf{R}) |^{2} - | \Delta_{\mathbf{p}+\mathbf{Q}_{0}+\mathbf{q}}(\mathbf{R})  |^{2}   ) \nonumber \\ 
&-& | \Delta_{\mathbf{p}}(\mathbf{R}) \varepsilon(\mathbf{p}+\mathbf{Q}_{0} + \mathbf{q})
- \Delta_{\mathbf{p}+\mathbf{Q}_{0}+\mathbf{q}}(\mathbf{R}) \varepsilon(\mathbf{p})|^{2}   ]  \nonumber \\ 
&+&  d_{3} b_{1} [ \varepsilon_{n}^{2} + I^{2} - \epsilon(\mathbf{p})^{2} 
- 2\varepsilon(\mathbf{p}) \varepsilon(\mathbf{p}+\mathbf{Q}_{0} + \mathbf{q})
- | \Delta_{\mathbf{p}+\mathbf{Q}_{0}+\mathbf{q}}(\mathbf{R})  |^{2}
- 2 \Delta_{\mathbf{p}}(\mathbf{R}) \Delta^{\ast}_{\mathbf{p}+\mathbf{Q}_{0}+\mathbf{q}}(\mathbf{R}) 
 ] 
\bigr]
, \nonumber \\ 
V^{(4)}_{m, \, 4}(\mathbf{p},\mathbf{q};{\bf R})
&=& \frac{2}{( a_{1}^{2}+b_{1}^{2} )( c_{4}^{2}+d_{4}^{2} )}
\bigl[ 
c_{3}[ 
( \varepsilon_{n}^{2}+I^{2} - \varepsilon(\mathbf{p})\varepsilon(\mathbf{p}+\mathbf{Q}_{0} + \mathbf{q}) 
- \Delta_{\mathbf{p}}(\mathbf{R}) \Delta^{\ast}_{\mathbf{p}+\mathbf{Q}_{0}+\mathbf{q}}(\mathbf{R})  ) \nonumber \\ 
&\times& (  \varepsilon_{n}^{2}-I^{2} - \varepsilon(\mathbf{p}) \varepsilon(\mathbf{p}+\mathbf{Q}_{0} + \mathbf{q}) 
- \Delta_{\mathbf{p}}(\mathbf{R}) \Delta^{\ast}_{\mathbf{p}+\mathbf{Q}_{0}+\mathbf{q}}(\mathbf{R}) ) 
- \varepsilon_{n}^{2} (   (\varepsilon(\mathbf{p}) + \varepsilon(\mathbf{p}+\mathbf{Q}_{0} + \mathbf{q}) )^{2} \nonumber \\ 
&+& | \Delta_{\mathbf{p}}(\mathbf{R}) + \Delta_{\mathbf{p}+\mathbf{Q}_{0}+\mathbf{q}}(\mathbf{R})  |^{2} ) 
+ I^{2} (\varepsilon(\mathbf{p})^{2} - \varepsilon(\mathbf{p}+\mathbf{Q}_{0} + \mathbf{q})^{2} 
+ | \Delta_{\mathbf{p}}(\mathbf{R}) |^{2} - | \Delta_{\mathbf{p}+\mathbf{Q}_{0}+\mathbf{q}}(\mathbf{R})  |^{2}   ) \nonumber \\ 
&+& | \Delta_{\mathbf{p}}(\mathbf{R}) \varepsilon(\mathbf{p}+\mathbf{Q}_{0} + \mathbf{q})
- \Delta_{\mathbf{p}+\mathbf{Q}_{0}+\mathbf{q}}(\mathbf{R}) \varepsilon(\mathbf{p}) |^{2}
] \nonumber \\ 
&+& d_{3} b_{1} [ \varepsilon_{n}^{2} + I^{2} - \varepsilon(\mathbf{p})^{2} 
- 2\varepsilon(\mathbf{p}) \varepsilon(\mathbf{p}+\mathbf{Q}_{0} + \mathbf{q})
- | \Delta_{\mathbf{p}}(\mathbf{R}) + \Delta_{\mathbf{p}+\mathbf{Q}_{0}+\mathbf{q}}(\mathbf{R}) |^{2}
+ | \Delta_{\mathbf{p}+\mathbf{Q}_{0}+\mathbf{q}}(\mathbf{R}) |^{2}
 ] 
\bigr]
, \nonumber \\ 
V^{(4)}_{m, \, 5}(\mathbf{p},\mathbf{q};{\bf R})
&=& \frac{2}{( a_{1}^{2}+b_{1}^{2} )( a_{2}^{2}+b_{1}^{2} )}
\bigl[ ( \varepsilon_{n}^{2}-I^{2}-\varepsilon(\mathbf{p})\varepsilon(\mathbf{p}+\mathbf{Q}_{0} + \mathbf{q}) 
- \Delta_{\mathbf{p}}(\mathbf{R}) \Delta^{\ast}_{\mathbf{p}+\mathbf{Q}_{0}+\mathbf{q}}(\mathbf{R}) )^{2} + b_{1}^{2}  \nonumber \\ 
&-& (\varepsilon_{n}^{2} + I^{2} )[ ( \varepsilon(\mathbf{p}) + \varepsilon(\mathbf{p}+\mathbf{Q}_{0} + \mathbf{q}) )^{2} 
+ |\Delta_{\mathbf{p}}(\mathbf{R}) + \Delta_{\mathbf{p}+\mathbf{Q}_{0}+\mathbf{q}}(\mathbf{R})  |^{2}  ] \nonumber \\
&-& |\Delta_{\mathbf{p}}(\mathbf{R}) \varepsilon(\mathbf{p}+\mathbf{Q}_{0} + \mathbf{q})
- \Delta_{\mathbf{p}+\mathbf{Q}_{0}+\mathbf{q}}(\mathbf{R}) \varepsilon(\mathbf{p}) |^{2}
\bigr]
\end{eqnarray*}
with
\begin{eqnarray}
a_{2} &=& \varepsilon_{n}^{2} + \varepsilon(\mathbf{p}+\mathbf{Q}_{0}+\mathbf{q})^{2} 
+ |\Delta_{\mathbf{p}+\mathbf{Q}_{0}+\mathbf{q}}(\mathbf{R})|^{2} - I^{2},  \nonumber \\ 
c_{1} &=& a_{1}a_{2} + b_{1}^{2}, \nonumber \\ 
d_{1} &=& b_{1} ( a_{2} - a_{1} ), \nonumber \\ 
c_{2} &=& a_{1}a_{2} - b_{1}^{2}, \nonumber \\ 
d_{2} &=& b_{1} ( a_{1} + a_{2} ), \nonumber \\ 
c_{3} &=& a_{1}^{2} - b_{1}^{2}, \nonumber \\ 
d_{3} &=& 2 a_{1} b_{1}, \nonumber \\ 
c_{4} &=& a_{2}^{2} - b_{1}^{2}, \nonumber \\ 
d_{4} &=& 2 a_{2} b_{1}. 
\end{eqnarray}

Here, we assume the Q2D Fermi surface distorted along the $c$ direction. It leads to an incommensurate wavevector of the AFM order directed to [1,1,0] in $\mathbf{H} \parallel ab$ case.

\end{document}